\documentclass[preprint,superscriptaddress,preprintnumbers,frontmatterverbose,amsmath,amssymb,aps,prd,floatfix,notitlepage,nofootinbib]{revtex4-2}

\usepackage{graphicx}
\usepackage{dcolumn}
\usepackage{bm}
\usepackage{hyperref}
\usepackage{orcidlink}
\usepackage{verbatim}
\usepackage{multirow}
\usepackage{enumitem}

\numberwithin{equation}{section}

\allowdisplaybreaks

\def \beq{\begin{equation}}
\def \eeq{\end{equation}}
\def \bea{\begin{eqnarray}}
\def \eea{\end{eqnarray}}
\def \bma{\begin{matrix}}
\def \ema{\end{matrix}}
\def \s{\sqrt{2}}
\def \({\left(}
\def \){\right)}
\def \[{\left[}
\def \]{\right]}
\def \nn{\nonumber}
\def \nl{\nn\\}
\def \hf{\frac{1}{2}}

\def\bra#1{\left\langle #1\right|}
\def\ket#1{\left| #1\right\rangle}

\newcommand\varpm{\mathbin{\vcenter{\hbox{%           %%%% For +/(-) 
  \oalign{\hfil$\scriptstyle+$\hfil\cr
          \noalign{\kern-.3ex}
          $\scriptscriptstyle({-})$\cr}%
}}}}
\newcommand\varmp{\mathbin{\vcenter{\hbox{%           %%%% For -/(+) 
  \oalign{\hfil$\scriptstyle-$\hfil\cr
          \noalign{\kern-.3ex}
          $\scriptscriptstyle({+})$\cr}%
}}}}

\newcommand{\f}[2]{\frac{#1}{#2}}    %% The fraction

\begin{document}
\count\footins = 1000

\title{\boldmath A global analysis of ${\bar B}\to DP$ and ${\bar B} \to {\bar D} P$ decays}

\author{Bhubanjyoti Bhattacharya\,\orcidlink{0000-0003-2238-321X}}
\email{bbhattach@ltu.edu}
\affiliation{Department of Natural Sciences, \\ Lawrence Technological University, Southfield, MI 48075, USA}

\author{Rida Boumris\,\orcidlink{0009-0001-0021-9593}}
\email{rida.boumris@umontreal.ca}
\affiliation{Physique des Particules, Universit\'e de Montr\'eal, 1375 Avenue Th\'er\`ese-Lavoie-Roux, Montr\'eal, QC, Canada  H2V 0B3}

\author{David London\,\orcidlink{0000-0002-4407-5624}}
\email{london@lps.umontreal.ca}
\affiliation{Physique des Particules, Universit\'e de Montr\'eal, 1375 Avenue Th\'er\`ese-Lavoie-Roux, Montr\'eal, QC, Canada  H2V 0B3}

\date{\today}

\begin{abstract}
Recently, a fit to the 10 measured branching ratios of ${\bar B} \to D P$ decays ($B = \{B^+, B^0, B_s^0\}$, $D = \{D^0, D^+, D_s^+\}$, $P \in \{ \pi, K\}$) was performed under the assumption of flavor SU(3) symmetry [SU(3)$_F$]. A terrible fit was found: there was a $> 5\sigma$ disagreement with the SU(3)$_F$ limit of the Standard Model [SM$_{\rm{SU(3)}_F}$]. It was also found that this discrepancy can be removed by adding $\sim 20\%$ SU(3)$_F$-breaking effects. Now, the ${\bar B} \to D^0 P$ amplitudes can contribute to other observables (direct and indirect CP asymmetries) if they interfere with the amplitudes for ${\bar B} \to {\bar D}^0 P$ decays. This happens if the $D^0$ and ${\bar D}^0$ decay to the same final state (e.g., $K^+ K^-$, $\pi^+ \pi^-$, $K^\pm \pi^\mp$, etc.). In this paper, we perform a combined fit of ${\bar B} \to D P$ and ${\bar B} \to {\bar D} P$ decays, including all observables. We again find a $> 5\sigma$ disagreement with the SM$_{\rm{SU(3)}_F}$, and we find that the data can be explained if $\sim 60\%$ SU(3)$_F$-breaking effects are included.
\end{abstract}

\preprint{UdeM-GPP-TH-26-312}

\maketitle
\clearpage

\section{Introduction}

Recently, $B\to PP$ decays ($B = \{B^+, B^0, B_s^0\}$, $P$ is a pseudoscalar meson) were examined under the assumption of flavor SU(3) symmetry [SU(3)$_F$]. In Ref.~\cite{Berthiaume:2023kmp}, $P$ was limited to $\{\pi, K\}$, while in Ref.~\cite{Bhattacharya:2025wcq}, the analysis was extended so that $P$ could be any of  $\{ \pi, K, \eta, \eta' \}$. Fits to the latest data were performed, and in both cases it was found that the full fit combining the $\Delta S=0$ and $\Delta S=1$ data is very poor: there is a significant ($\sim 4\sigma$) disagreement with the SU(3)$_F$ limit of the standard model (SM$_{\rm{SU(3)}_F}$). This discrepancy can be removed by adding SU(3)$_F$-breaking effects, but 1000\% SU(3)$_F$ breaking is required. This is much larger than the $\sim 30\%$
SU(3)$_F$ breaking expected in the SM.

A similar analysis of ${\bar B} \to D P$ decays ($D = \{D^0, D^+, D_s^+\}$, $P = \{ \pi, K\}$) was performed in Ref.~\cite{Davies:2024vmv}. There it was also found that the combined fit of all $\Delta S=0$ and $\Delta S=1$ data shows a sizeable ($> 5\sigma$) disagreement with SM$_{\rm{SU(3)}_F}$. However, in this case it was shown that this discrepancy can be removed with the addition of $\sim 20\%$ SU(3)$_F$-breaking effects.

This study used only the branching ratios of ${\bar B} \to D P$ decays. However, the amplitudes for ${\bar B} \to D^0 P$ decays can interfere with the amplitudes for ${\bar B} \to {\bar D}^0 P$ decays if the $D^0$ and ${\bar D}^0$ decay to the same final state. There are two possibilities: both $D^0$ and ${\bar D}^0$ decay to (i)  a CP-eigenstate final state such as $K^+ K^-$ or $\pi^+ \pi^-$ \cite{Gronau:1990ra, Gronau:1991dp}, or (ii) a non-CP-eigenstate final state such as $K^+ \pi^-$. In the second case, the $D^0$ decay is doubly-Cabibbo-suppressed, while the ${\bar D}^0$ decay is Cabibbo-favoured. The CP phase $\gamma$ is generally extracted using methods that rely on this interference. 

The point is that the analysis of Ref.~\cite{Davies:2024vmv} could be extended to also include observables related to the interference of the ${\bar B} \to D^0 P$ and ${\bar B} \to {\bar D}^0 P$ amplitudes. Of course, this would also require adding the branching ratios of the ${\bar B} \to {\bar D} P$ decays. In addition, although it was shown that the discrepancy with SM$_{\rm{SU(3)}_F}$ can be removed with the addition of $\sim 20\%$ SU(3)$_F$-breaking effects, it was never explicitly shown that SU(3)$_F$ is not respected, i.e., there was no comparison of the $\Delta S=0$ and $\Delta S=1$ decays.

In the present paper, we modify the ${\bar B} \to D P$ analysis of Ref.~\cite{Davies:2024vmv} to address all of the above points. As we will see, the conclusions are changed a bit, but not enormously: when we perform a combined fit of  ${\bar B}\to DP$ and ${\bar B}\to{\bar D}P$ decays, we again find a very poor fit in the SU(3)$_F$ limit. The disagreement with the SM$_{\rm{SU(3)}_F}$ can be removed when 60\% SU(3)$_F$-breaking effects are added.

We begin in Sec.~II by discussing the formalism used to describe the ${\bar B}\to DP$ and ${\bar B}\to{\bar D}P$ decay amplitudes. We discuss the decomposition of the amplitudes in terms of reduced matrix elements and topological diagrams, both in the SU(3)$_F$ limit and with SU(3)$_F$-breaking corrections. (The equivalence of reduced matrix elements and diagrams is shown in Appendix \ref{RME_Diag_Equiv}.) In Sec.~III we examine the 12 ${\bar B}\to DP$ decays and 16 ${\bar B}\to{\bar D}P$ decay amplitudes. The observables used in our analysis are discussed in Sec.~IV. Sec.~V contains the results of the fits to the data. These include the individual ${\bar B}\to DP$ and ${\bar B}\to{\bar D}P$ fits, as well as the combined fit. We conclude in Sec.~VI.

\section{Formalism} 
\label{sec:formalism}

Our analysis involves the study of ${\bar B}\to DP$ and ${\bar B}\to{\bar D}P$ decays, $P \in \{\pi, K\}$, both in the limit of flavour SU(3) symmetry, and including SU(3)$_F$-breaking effects. We use the Wigner-Eckart theorem to express all decay amplitudes in terms of SU(3)$_F$ reduced matrix elements (RMEs) or, equivalently, topological diagrams. These RMEs/diagrams are then used as the unknown parameters in a fit to the data. In this way we can ascertain how well the SM explains the measurements, first in the SU(3)$_F$ limit, and then with the inclusion of SU(3)$_F$ breaking.

In this Section, we describe the formalism of our analysis. In particular, for both ${\bar B}\to DP$ and ${\bar B}\to{\bar D}P$ decays we work out the RMEs in the SU(3)$_F$ limit, as well as those that are produced when SU(3)$_F$ breaking is introduced. Similarly, we present the various diagrams that are found in the amplitudes, both in the SU(3)$_F$ limit and with SU(3)$_F$ breaking.

\subsection{Reduced Matrix Elements}

\subsubsection{\boldmath The SU(3)$_F$ Limit}

The amplitudes for the two-body ${\bar B}$ decays we consider here are of the form 
%$\langle \overset{(-)}{D} P|{\cal H}_W|{\bar B}\rangle$, 
$\langle f |{\cal H}_W|{\bar B}\rangle$,
where $f = DP$ or ${\bar D}P$ and ${\cal H}_W$ is the weak transition Hamiltonian. In order to apply the Wigner-Eckart theorem and obtain the RMEs, we must determine how the three pieces ${\bar B}$, 
%$\overset{(-)}{D} P$ 
$f$ and ${\cal H}_W$ transform under SU(3)$_F$. The initial state, ${\bar B} = (B^-, {\bar B}^0, {\bar B}^0_s)$, is an antitriplet, $\mathbf{3^*}$. The $D = (D^0, D^+, D_s^+)$ is also a $\mathbf{3^*}$, while $P = \{ \pi^\pm, \pi^0, K^\pm, K^0, {\bar K}^0 \}$ transforms as an octet, $\mathbf{8}$. The $DP$ final state therefore transforms under SU(3)$_F$ as $\mathbf{3^*} \otimes \mathbf{8}=\mathbf{15^*} \oplus \mathbf{6}\oplus \mathbf{3^*}$, while ${\bar D}P$ transforms as $\mathbf{3} \otimes \mathbf{8} = \mathbf{15} \oplus \mathbf{6^*} \oplus \mathbf{3}$. 

The effective Hamiltonian for these transitions is given by \cite{Gronau:1995hm}
\begin{equation}\label{eq:H}
    \mathcal{H}_W = \frac{G_F}{\sqrt{2}}\sum_{q=d,s} \Big \{ \lambda_1 Q_1^q + \lambda_2 Q_2^q\Big \} ~.
\end{equation}
Here $\lambda_1 = V_{cb}^{}V_{uq}^*$ and $\lambda_2 = V_{ub}^{}V_{cq}^*$, where the $V_{ij}^{}$ are elements of the Cabibbo-Kobayashi-Maskawa (CKM) matrix. The operators $Q_1^q$ and $Q_2^q$ are 
\begin{equation}
Q_1^q  = (b\bar c)_{V-A} (u \bar q)_{V-A} ~~,~~~~ 
Q_2^q  = (b \bar u)_{V-A} (c \bar q)_{V-A} ~.
\end{equation}
$Q_1^q$ transforms as an $\mathbf{8}$, while $Q_2^q$ transforms as $\mathbf{3^*}\oplus\mathbf{6}$. It is useful to keep track of the isospin and hypercharge quantum numbers of these operators. $Q_1^q$ and $Q_2^q$ can be expressed as
\begin{eqnarray}
    Q_1^{d(s)} &=&  - ~\mathbf{8}_{I=1(\hf)}^{Y=0(-1)} ~, \nn\\
    Q_2^{d(s)} &=&  \varmp~\frac{1}{\sqrt{2}} ~ \mathbf{3^*}_{Y=\frac{2}{3} (-\f{1}{3})}^{I=0(\hf)} + \frac{1}{\sqrt{2}} ~ \mathbf{6}_{Y=\frac{2}{3}(-\f{1}{3})}^{I=1(\hf)} ~.
\end{eqnarray}
The information about the quantum numbers is necessary to compute the SU(3)$_F$ Clebsch-Gordan coefficients. It will also be important when we introduce SU(3)$_F$-breaking effects, in which case $\Delta S=0$ and $\Delta S=1$ decays will generally be affected differently.

The ${\bar B}\to DP$ decay amplitudes therefore involve the following three RMEs:
\begin{equation}\label{eq:RMEs}
        O_{\bar{15}}  = \big < \mathbf{15^*} \big| \big| \, \mathbf{8} \, \big | \big|  \mathbf{3^*}\big> ~~,~~~~
        O_{6}  = \big <  \mathbf{6} \big| \big| \,  \mathbf{8} \, \big | \big|  \mathbf{3^*}\big> ~~,~~~~
        O_{\bar 3}  = \big <  \mathbf{3^*} \big| \big| \,  \mathbf{8} \, \big | \big|  \mathbf{3^*}\big> ~,
\end{equation}
while the following four RMEs appear in the ${\bar B}\to {\bar D}P$ decay amplitudes:
\beq
 S_{15}  = \big \langle \mathbf{15} || \ \mathbf{6}\ ||  \mathbf{3^*}\big \rangle ~,~~
 S_{3}  = \big \langle \mathbf{3} || \ \mathbf{6}\ ||  \mathbf{3^*}\big \rangle ~,~~
 T_{\bar 6}  = \big \langle \mathbf{6^*} || \ \mathbf{3^*}\ ||  \mathbf{3^*}\big \rangle ~,~~
 T_{3}  = \big \langle \mathbf{3} || \ \mathbf{3^*}\ ||  \mathbf{3^*}\big \rangle ~.
\eeq
In the fits, each RME is treated as a complex parameter. The global fit, which combines both ${\bar B}\to DP$ and ${\bar B}\to {\bar D}P$ decays, therefore involves seven RMEs. This corresponds to 13 real parameters (seven magnitudes and six relative strong phases). 

\subsubsection{SU(3)$_F$-Breaking Corrections}

Because the mass of the $s$ quark is larger than those of the $u$ and $d$ quarks, SU(3)$_F$ is not an exact symmetry -- there are SU(3)$_F$-breaking corrections due to this mass difference. The operator representing the quark masses can be expressed as follows:
\bea
   Q_M &=& m_u\,u\bar u + m_d\,d\bar d + m_s\,s\bar s \nn\\
   &=& \frac13 (m_u + m_d + m_s)(u\bar u + d\bar d + s\bar s) + \frac12 (m_u - m_d)(u\bar u - d\bar d) \nn\\
   && \hskip4truemm
   +~\frac16 (m_u + m_d - 2m_s)(u\bar u + d\bar d - 2 s\bar s) ~.
   \label{eq:Qm}
\eea
The first term above is an SU(3)$_F$ singlet. The second term represents isospin breaking via $(m_u - m_d)$. The third term accounts for SU(3)$_F$ breaking through $(m_u + m_d - 2m_s)$. Note that both the second and third terms transform as different components of an $\mathbf{8}$ of SU(3)$_F$. 

As a first approximation, we ignore isospin breaking and focus only on SU(3)$_F$-breaking effects. We treat the third term in Eq.~(\ref{eq:Qm}) as a spurion field. Under SU(3)$_F$, this operator transforms as
\begin{equation}
\label{eq:Q_breaking}
   Q_{\rm brk} = u\bar u + d\bar d - 2s\bar s = -\sqrt{6} ~ \mathbf{8}^{Y=0}_{I=0} ~.
\end{equation}
The first-order SU(3)$_F$-breaking correction is found by
multiplying each of the initial state, the Hamiltonian, and the final state by $Q_{\rm brk}$: 
\begin{equation}
\label{eq:Break_correction}
    \epsilon^{(1)} = \bra{f} \mathcal{H_W}\ket{Q_{\rm brk}\otimes B} + \bra{f} \mathcal{H_W}\otimes Q_{\rm brk}\ket{B}  +  \bra{Q_{\rm brk} \otimes f} \mathcal{H_W}\ket{B} ~.
\end{equation}
There are therefore three pieces that contribute to first-order SU(3)$_F$ breaking. 

Before computing these contributions, it is worthwhile discussing their physics interpretation. For the initial state, the correction $Q_{\rm brk}\otimes B$ corresponds to the fact that the masses of the 
$(B^-, {\bar B}^0)$ and ${\bar B}^0_s$ mesons are different. Similarly, the correction to the final state, $Q_{\rm brk} \otimes f$, 
is due to the fact that the masses of $(D^0, D^+)$ and $D_s^+$, and of $\pi$ and $K$, are different. However, this is already accounted for in the fits by including a phase-space factor in the definition of the branching ratios that involves the physical masses of the particles in the decay [see Eq.~(\ref{eq:Br})]. With this, it is no longer necessary to account for the SU(3)$_F$-breaking corrections to the initial and final states.

We therefore consider only the correction to the Hamiltonian, $\mathcal{H_W}\otimes Q_{\rm brk}$. This will reflect the fact that, because of SU(3)$_F$ breaking, $b \to s$ transitions may differ from $b \to d$ transitions. The inclusion of this correction will generate new RMEs to be added to the previous SU(3)$_F$-invariant RMEs. 

For ${\bar B}\to DP$, the correction to $\mathcal{H_W}$ transforms under SU(3)$_F$ as follows: $\mathbf{8}\otimes\mathbf{8} = \mathbf{27}\oplus \mathbf{10^*}\oplus \mathbf{10}\oplus \mathbf{8_1}\oplus \mathbf{8_2}\oplus\mathbf{1}$. More specifically, when we multiply $Q_1$ by $Q_{\rm brk}$, we find
\bea
    Q_1^{(d)} \otimes Q_{\rm brk}  &=&  \mathbf{10}^{Y=0}_{I=1}  +  \mathbf{10^*}^{Y=0}_{I=1}  +  \mathbf{27}^{Y=0}_{I=1}  + \mathbf{8_1}^{Y=0}_{I=1}  ~, \nn\\
     Q_1^{(s)} \otimes Q_{\rm brk}  &=&   \mathbf{10}^{Y=-1}_{I=\frac{1}{2}}  +  \sqrt{\frac{3}{2}}\mathbf{27}^{Y=-1}_{I=\frac{1}{2}}  - \frac{1}{2}\mathbf{8_1}^{Y=-1}_{I=\frac{1}{2}}  -  \mathbf{8_2}^{Y=-1}_{I=\frac{1}{2}} ~. 
\eea
These lead to the following 9 SU(3)$_F$-breaking RMEs:
\bea
\label{eq:RME_breaking}
&        O^1_{\bar{15}} =  \big < \mathbf{15^*} \big|\big| \, \mathbf{8_1} \, \big|\big| \mathbf{3^*}\big > ~,~~
        O^1_{6} = \big < \mathbf{6} \big|\big| \, \mathbf{8_1} \, \big|\big| \mathbf{3^*}\big > ~,~~
        O^1_{\bar{3}} = \big < \mathbf{3^*} \big|\big| \, \mathbf{8_1} \, \big|\big| \mathbf{3^*}\big > ~, & \nn\\
&        O^2_{\bar{15}} = \big < \mathbf{15^*} \big|\big| \, \mathbf{8_2} \, \big|\big| 3^*\big > ~,~~
        O^2_{6}  = \big < \mathbf{6} \big|\big| \, \mathbf{8_2} \, \big|\big| \mathbf{3^*}\big > ~,~~
        O^2_{\bar{3}} = \big < \mathbf{3^*} \big|\big| \, \mathbf{8_2} \, \big|\big| \mathbf{3^*}\big > ~, & \nn\\
&        W_{\bar{15}} = \big < \mathbf{15^*} \big|\big| \, \mathbf{10^*} \, \big|\big| \mathbf{3^*}\big > ~,~~
        Z_{\bar{15}} = \big < \mathbf{15^*} \big|\big| \,\mathbf{27} \, \big|\big| \mathbf{3^*}\big > ~,~~
        V_{6}  =  \big <\mathbf{6} \big|\big| \, \mathbf{10} \, \big|\big| \mathbf{3^*}\big > ~. &
\eea
When one includes the three SU(3)$_F$-invariant RMEs, this means that the ${\bar B}\to DP$ decay amplitudes are expressed in terms of 12 total RMEs.

${\bar B}\to {\bar D}P$ decays are analyzed in a similar manner. The SU(3)$_F$-breaking corrections to $\mathcal{H_W}$ transform as $\mathbf{8}\otimes (\mathbf{3^*} \oplus \mathbf{6}) = \mathbf{15^*_1}\oplus \mathbf{6}_1 \oplus \mathbf{3^*_1} \oplus \mathbf{24^*} \oplus \mathbf{15^*_2} \oplus \mathbf{6_2} \oplus \mathbf{3^*_2}$. More explicitly,
\begin{eqnarray}
    Q_2^{(d)} \otimes Q_{\rm brk} &=& \mathbf{15^*_1}^{Y=\frac{2}{3}}_{I=0} +  \mathbf{15^*_2}^{Y=\frac{2}{3}}_{I=1} - \mathbf{24^*}^{Y=\frac{2}{3}}_{I=1} - \mathbf{3^*_1}^{Y=\frac{2}{3}}_{I=0} + \mathbf{6_1}^{Y=\frac{2}{3}}_{I=1} ~, \nn\\
    Q_2^{(s)} \otimes Q_{\rm brk}   &=&   - ~ \frac{\sqrt{3}}{2}\mathbf{15^*_1}^{Y=\frac{-1}{3}}_{I=\frac{1}{2}} + \frac{\sqrt{3}}{2\sqrt{2}}\mathbf{15^*_2}^{Y=\frac{-1}{3}}_{I=\frac{1}{2}}  - \sqrt{\frac{3}{2}}\mathbf{24^*}^{Y=\frac{-1}{3}}_{I=\frac{1}{2}}  \nl 
    && \hspace{1truecm} -~\frac{1}{2}\mathbf{3^*_1}^{Y=\frac{-1}{3}}_{I=\frac{1}{2}}  + \mathbf{3^*_2}^{Y=\frac{-1}{3}}_{I=\frac{1}{2}}  + \mathbf{6_2}^{Y=\frac{-1}{3}}_{I=\frac{1}{2}} -\frac{1}{2} \mathbf{6_1}^{Y=\frac{-1}{3}}_{I=\frac{1}{2}} ~.   
\end{eqnarray}
These produce the following 13 SU(3)$_F$-breaking RMEs: 
\bea
       & M^1_{15} = \big < \mathbf{15} \big|\big| \, \mathbf{{15^*_1}} \, \big|\big| \mathbf{3^*}\big > ~,~~
        M^2_{15} = \big < \mathbf{15} \big|\big| \, \mathbf{{15^*_1}} \, \big|\big| \mathbf{3^*}\big > ~,~~
        M^1_{\bar{6}} = \big < \mathbf{6^*} \big|\big| \, \mathbf{{15^*_1}} \, \big|\big| \mathbf{3^*}\big > ~, & \nn\\
       & M^2_{\bar{6}} = \big < \mathbf{6^*} \big|\big| \, \mathbf{{15^*_2}} \, \big|\big| \mathbf{3^*}\big > ~,~~
       R_{15} = \big < \mathbf{15} \big|\big| \, \mathbf{{24^*}} \, \big|\big| \mathbf{3^*}\big > ~, & \\
        & T_{3}^1 = \big < \mathbf{3}\big|\big| \, \mathbf{{3^*_1}} \, \big|\big| \mathbf{3^*}\big > ~,~~
        T_{3}^2 = \big < \mathbf{3} \big|\big| \, \mathbf{{3^*_2}} \, \big|\big| \mathbf{3^*}\big > ~,~~
        T_{\bar{6}}^1 = \big < \mathbf{6^*}\big|\big| \, \mathbf{{3^*_1}} \, \big|\big| \mathbf{3^*}\big > ~,~~
         T_{\bar{6}}^2 = \big < \mathbf{6^*} \big|\big| \, \mathbf{{3^*_2}} \, \big|\big| \mathbf{3^*}\big > ~, & \nn\\
        & S_{15}^1 = \big < \mathbf{15} \big|\big| \, \mathbf{6_1} \, \big|\big| \mathbf{3^*}\big > ~,~~
        S_{15}^2 = \big < \mathbf{15} \big|\big| \, \mathbf{6_2} \, \big|\big| \mathbf{3^*}\big > ~,~~
        S_{3}^1 = \big < \mathbf{3} \big|\big| \, \mathbf{6_1} \, \big|\big| \mathbf{3^*}\big > ~,~~
        S_{3}^2 = \big < \mathbf{3} \big|\big| \, \mathbf{6_2} \, \big|\big| \mathbf{3^*} \big > ~, & \nn
\eea
Given that there are also four SU(3)$_F$-invariant RMEs, the ${\bar B}\to {\bar D}P$ decay amplitudes are expressed in terms of a total of 17 RMEs.

\subsection{Topological Diagrams}

\subsubsection{\boldmath The SU(3)$_F$ Limit}

Alternatively, the decay amplitudes can be parametrized in terms of topological diagrams. The diagrams that contribute to charmless $B \to PP$ decays are described in Refs.~\cite{Gronau:1994rj, Gronau:1995hn}. The diagrams that contribute to ${\bar B}\to DP$ and ${\bar B}\to {\bar D}P$ decays are similar.

For ${\bar B}\to DP$, there are three contributing diagrams: a tree diagram $T$, a colour-suppressed tree diagram $C$, and an exchange diagram $E$. For ${\bar B}\to {\bar D}P$, in addition to these three diagrams, the annihilation diagram $A$ also contributes. In Ref.~\cite{Gronau:1994rj}, it is argued that $|E/T|$ and $|A/T|$ are both $\ll 1$. In our fits, we make no assumptions about the relative sizes of diagrams, but these expectations are borne out by the data: the branching ratios of decays whose amplitudes are purely $E$ and/or $A$ are much smaller than the branching ratios of decays that proceed via $T$.

Since the operators $Q_1$ and $Q_2$ are not the same, the diagrams for ${\bar B} \to DP$ and ${\bar B} \to {\bar D}P$ decays are also different. For this reason, we write the ${\bar B} \to {\bar D}P$ diagrams with primes.

Note that, for each decay, we have the same number of diagrams and RMEs (three for ${\bar B}\to DP$ and four for ${\bar B}\to {\bar D}P$). This suggests an equivalence between the two descriptions. This is discussed in the next section and is shown explicitly in Appendix \ref{RME_Diag_Equiv}.

\subsubsection{SU(3)$_F$-Breaking Corrections}

The SU(3)$_F$-breaking diagrams that contribute to charmless $B \to PP$ decays are described in Ref.~\cite{Gronau:1995hm}. The procedure is simple: for a given diagram ${\cal D}$, new diagrams are defined by putting an `X' on each $s$-quark line. This `X' represents the fact that there is a mass difference between the $s$ quark and $u$/$d$ quarks. 

We adopt a similar procedure here, defining the SU(3)$_F$-breaking diagrams by placing an `X' on an $s$-quark line. For ${\cal D} = T,C$, ${\cal D}_1$ is the diagram in which an $s$ quark is connected to a $W$ vertex, while ${\cal D}_2$ is the diagram in which the $s$ quark is the spectator. For ${\cal D} = E,A$, ${\cal D}_1$ is again the diagram in which an $s$ quark is connected to a $W$ vertex, while ${\cal D}_2$ is the diagram in which an $s {\bar s}$ quark pair is popped from the vacuum.  

For ${\bar B}\to DP$ decays, there are three SU(3)$_F$-invariant diagrams and six SU(3)$_F$-breaking diagrams, for a total of nine. For ${\bar B}\to {\bar D}P$ decays, there are twelve diagrams in total, four SU(3)$_F$-invariant and eight SU(3)$_F$-breaking.

\section{\boldmath ${\bar B}\to DP$ and ${\bar B}\to {\bar D}P$ Decay Amplitudes}
\label{Sec: amplitudes}

There are 12 ${\bar B}\to DP$ decays and 16 ${\bar B}\to{\bar D}P$ decays, listed in Table \ref{tab:List_decays_B_DP}. As shown in the previous section, the amplitudes for these decays can be expressed in terms of RMEs or diagrams. For the purposes of a fit to the data, diagrams are more useful. The reason is that each SU(3)$_F$-breaking diagram is associated with an SU(3)$_F$-invariant diagram. This allows us to quantify the size of SU(3)$_F$ breaking\footnote{In our analysis, the size of the SU(3)$_F$-breaking diagrams is obtained from the fits. In Ref.~\cite{Bordone:2020gao}, the size of the SU(3)$_F$-breaking contributions to ${\bar B}^0_{(s)} \to D^+_{(s)} \{\pi^-, K^-\}$ is computed using QCD factorization. This provides complementary information.}. For example, if $|T_1/T|$ is found to be 0.3, we can say that this corresponds to 30\% SU(3)$_F$ breaking. This cannot be done with RMEs, since the SU(3)$_F$-breaking RMEs are unrelated to the SU(3)$_F$-invariant RMEs.

\begin{table}[h!]
    \centering
    \begin{tabular}{c c|c c}
    \multicolumn{4}{c}{${\bar B}\to DP$}\\
    \hline
        \multicolumn{2}{c|}{$\Delta S=0$ }& \multicolumn{2}{|c}{$\Delta S=1$ } \\
        \hline \hline
        $B^- \to$&$ D^0\pi^-$ &  $B^- \to $&$ D^0K^-$\\
        \hline
        $ \bar{B}^0 \to$&$ D^+\pi^-$ & $ \bar{B}^0_s\to$&$ D^+_s K^-$ \\
        $ \bar{B}^0 \to$&$ D^+_sK^-$ & $ \bar{B}^0_s\to $&$D^+\pi^-$ \\
        $ \bar{B}^0 \to $&$D^0\pi^0$  &  $ \bar{B}^0_s \to $ & $D^0 \pi^0$ \\
         \hline
         $ \bar{B}^0_s \to$&$ D^0K^0$ &  $ \bar{B}^0 \to $&$D^0  \bar{K}^0$\\
         $\bar{B_s^0} \to$&$ D_s^+ \pi^-$  & $ \bar{B}^0 \to $&$ D^+ K^-$ \\
          \hline \hline
    \end{tabular}
~~~~~~~~~~~~
    \begin{tabular}{c c|c c}
    \multicolumn{4}{c}{$\bar{B}\to \bar DP$}\\
    \hline
        \multicolumn{2}{c|}{$\Delta S=0$ }& \multicolumn{2}{|c}{$\Delta S=1$ } \\
        \hline \hline
        $B^- \to$&$ \bar D^0 \pi^- $ &  $B^- \to $&$ \bar D^0 K^-  $\\
        $B^- \to$&$ D^-_s K^0 $ &  $B^- \to $&$ D^- \bar K^0 $  \\
        $B^- \to$&$D^- \pi^0 $ &  $B^- \to$ & $ D^-_s \pi^0 $ \\
        
        \hline
        $ \bar{B}^0 \to$&$ D^- \pi^+ $ & $ \bar{B}^0_s\to$&$   D^-_s K^+ $ \\
        $ \bar{B}^0 \to $&$ D_s^- K^+ $ & $ \bar{B}^0_s \to $&$ D^- \pi^+ $   \\
        $ \bar{B}^0 \to $&$\bar D^0 \pi^0 $ &  $ \bar{B}^0_s \to $&$ \bar D^0 \pi^0$  \\
         \hline
         $ \bar{B}^0_s \to$&$  D^- K^+ $ &  $ \bar{B}^0 \to $ & $  D_s^- \pi^+ $\\
         $\bar{B_s^0} \to$&$ \bar D^0 K^0 $  & $ \bar{B}^0 \to $&$  \bar D^0 \bar K^0 $ \\
         
          \hline \hline
    \end{tabular}
  \caption{List of $\bar{B}\to DP$ decays (left table) and $\bar{B}\to {\bar D}P$ decays (right table). \\}
    \label{tab:List_decays_B_DP}
\end{table}

For this reason, from now on, we use only diagrams, both SU(3)$_F$ invariant and SU(3)$_F$ breaking, in our analysis. This applies to the decomposition of the ${\bar B}\to DP$ and ${\bar B}\to{\bar D}P$ decay smplitudes in this section, and to the fits of Sec.~V. We explicitly demonstrate the equivalence of RMEs and diagrams in Appendix \ref{RME_Diag_Equiv}.

We begin with ${\bar B}\to DP$ decays. Here there are three SU(3)$_F$-invariant diagrams, $T$, $C$ and $E$, and six SU(3)$_F$-breaking diagrams, $T_{1,2}$, $C_{1,2}$ and $E_{1,2}$, for a total of nine diagrams. However, when the amplitudes are expressed in terms of these diagrams, we find that only eight linear combinations appear. In other words, we can eliminate one diagram, which we choose to be $E_2$, and define eight effective diagrams:
\bea
&        \tilde{T} =  T - E_2 ~,~~
        \tilde{C}  = C + E_2 ~,~~
        \tilde{E}  = E + E_2 ~,~~
        \tilde{T}_1  = T_1 + E_2 ~, & \nn\\
&        \tilde{T}_2 =  T_2 + E_2 ~,~~
        \tilde{C}_1  = C_1 - E_2 ~,~~
        \tilde{C}_2  = C_2 - E_2 ~,~~
        \tilde{E_1}  = E_1 - E_2 ~. &
\eea

Note that $\tilde{T}$, $\tilde{C}$ and $\tilde{E}$ are the SU(3)$_F$-invariant diagrams with an SU(3)$_F$-breaking correction, while the other five effective diagrams are combinations of SU(3)$_F$-breaking diagrams. Above we said that $|T_1/T|$ provides a good measure of the size of SU(3)$_F$ breaking. Consider ${\tilde T}_1/{\tilde T}$:
\beq
\frac{{\tilde T}_1}{{\tilde T}} = \frac{T_1 + E_2}{T - E_2} \simeq
\frac{T_1}{T} \left[ 1 + \frac{E_2}{T} + \frac{E_2}{T_1} \right] ~.
\eeq
In Ref.~\cite{Gronau:1994rj}, it is argued that $|E/T|$ is $\ll 1$. This implies that $|E_2/T|$ is completely negligible, and also suggests that $|E_2/T_1|$ is small. The point is that, if the value of $|T_1/T|$ indicates the size of SU(3)$_F$ breaking, the value of $|{\tilde T}_1/{\tilde T}|$ does as well.

The decomposition of all 12 ${\bar B}\to DP$ decay amplitudes in terms of the eight effective diagrams is given in Table \ref{tab:Diag_contrib_DC1}.

\begin{table}[h!]
 \begin{tabular}{c|ccc|ccccc}
     {Decay mode ~}  &  ~~$\tilde{T}$ ~~ &  ~~$\tilde{C}$ ~~ &  ~~$\tilde{E}$ ~~ &  ~~$\tilde{T}_1$ ~~ &  ~~$\tilde{T}_2$ ~~ &  ~~$\tilde{C}_1$ ~~ &  ~~$\tilde{C}_2$ ~~ &  ~~$\tilde{E}_1$ ~~ \\
     \hline \hline
       $B^- \to D^0 \pi^-$ ~  & $-1$~~ & $-1$~~ & & & & & &   \\
       \hline
       $\bar{B}^0_s \to D^0 K^0$ ~ && $-1$~~ & & & & & $-1$~~ & \\
       \hline
       $\bar{B}^0_s \to D_s^+ \pi^-$  ~  & $-1$~~ & & & & $-1$~~  & & & \\
       \hline
       $\bar{B}^0 \to D^+\pi^-$  ~ & $-1$~~ & & $-1$~~ & & & & &  \\
       \hline
       $\bar{B}^0 \to D^0 \pi^0$  ~ & & $-\frac{1}{\sqrt{2}}$~~ & $\frac{1}{\sqrt{2}}$~~ & & & & & \\
       \hline
       $\bar{B}^0 \to D^+_s K^-$  ~ & & & $-1$~~ & & & & & \\
       \hline \hline
        $B^- \to D^0 K^-$ ~  & $-1$~~ & $-1$~~ & & $-1$~~ & & $-1$~~ & & \\
        \hline
        $\bar{B}^0 \to D^+ K^-$ ~ & $-1$~~ & & & $-1$~~ & & & & \\
        \hline
        $\bar{B}^0 \to D^0 \bar{K}^0$ ~ & & $-1$~~ & & & & $-1$~~ & &  \\
        \hline
        $\bar{B}^0_s \to D^+\pi^-$ ~  & & & $-1$~~ & & & & & $-1$~~  \\
        \hline
        $\bar{B}^0_s \to D^0 \pi^0$ ~ & & & $\frac{1}{\sqrt{2}}$~~ & & & & & $\frac{1}{\sqrt{2}}$~~ \\
        \hline
        $\bar{B}^0_s \to D^+_s K^-$ ~  & $-1$~~ & & $-1$~~ & $-1$~~ & $-1$~~ & & & $-1$~~  \\
       \hline
 \end{tabular}
 \caption{Decomposition of the ${\bar B}\to DP$ decay amplitudes in terms of effective diagrams. The first six amplitudes are $\Delta S =0$ ($\propto V_{cb}^{}V_{ud}^*$), while the last six amplitudes are $\Delta S =1$ ($\propto V_{cb}^{}V_{us}^*$). \\}
\label{tab:Diag_contrib_DC1}
\end{table}

We now turn to ${\bar B}\to {\bar D}P$ decays, where there are four SU(3)$_F$-invariant diagrams, $T'$, $C'$, $E'$ and $A'$, and eight SU(3)$_F$-breaking diagrams, $T'_{1,2}$, $C'_{1,2}$, $E'_{1,2}$ and $A'_{1,2}$, for a total of 12 diagrams. But here we find that only 11 linear combinations of diagrams appear in the expressions for the amplitudes. Eliminating $A'_2$, we define 11 effective diagrams:
\bea
&        \tilde{T}' =  T' + A'_2 ~,~~
        \tilde{C}'  =  C' - A'_2 ~,~~
        \tilde{E}'  =  E' - A'_2 ~,~~
        \tilde{A}'  =  A' + A'_2 ~, & \nn\\
&        \tilde{T}'_1 =  T'_1 - A'_2 ~,~~
        \tilde{T}'_2  =  T'_2 - A'_2 ~,~~
        \tilde{C}'_1  =  C'_1 + A'_2 ~,~~
        \tilde{C}'_2 =  C'_2 + A'_2 ~, & \nn\\
&        \tilde{E}'_1  =  E'_1 + A'_2 ~,~~
        \tilde{E}'_2  =  E'_2 + A'_2 ~,~~
        \tilde{A}'_1  =  A'_1 - A'_2 ~. &
\eea
Similar to what was found with ${\bar B}\to DP$ decays, here $\tilde{T}'$, $\tilde{C}'$, $\tilde{E}'$ and $\tilde{A}'$ are the SU(3)$_F$-invariant diagrams with an SU(3)$_F$-breaking correction, and the other seven effective diagrams are purely combinations of SU(3)$_F$-breaking diagrams.

Table \ref{tab:Diag_contrib_DC2} shows the decomposition of all 16 ${\bar B}\to {\bar D}P$ decay amplitudes in terms of the 11 effective diagrams.

\begin{table}[h!!]
     \centering
     \begin{tabular}{c|cccc|ccccccc}
     Decay mode~ & ~~$\tilde{T}'$~~ & ~~$\tilde{C}'$~~ & ~~$\tilde{E}'$~~ & ~~$\tilde{A}'$~~ & ~~$\tilde{T}'_1$~~ & ~~$\tilde{T}'_2$~~ & ~~$\tilde{C}'_1$~~ & ~~$\tilde{C}'_2$~~ & ~~$\tilde{E}'_1$~~ & ~~$\tilde{E}'_2$~~ & ~~$\tilde{A}'_1$~~ \\ 
     \hline \hline
       $B^- \to \bar{D}^0 \pi^-$ &   & $-1$ &  & $-1$ & & & & & & &   \\ 
       \hline
       $B^- \to D^- \pi^0$     & $-\frac{1}{\sqrt{2}}$ & & & $\frac{1}{\sqrt{2}}$ & & & & & & & \\ 
\hline
$B^- \to D^-_s K^0$    & & & & $1$ & & & & & & &  \\ 
       \hline 
       $ \bar{B}^0_s \to D^- K^+$ & $-1$ & & & & & $-1$ & & & & &  \\
\hline
$\bar{B}^0_s \to \bar{D}^0 K^0$ & & $-1$ & & & & & & $-1$ & & & \\ 
        \hline 
        $ \bar{B}^0 \to D^- \pi^+$ & $-1$ & & $-1$ & & & & & & & &\\
\hline
$\bar{B}^0 \to D_s^- K^+$  & & & $-1$ & &  &  & & & & $-1$ & \\
\hline
$\bar{B}^0 \to \bar{D}^0 \pi^0$  & & $-\frac{1}{\sqrt{2}}$ & $ \frac{1}{\sqrt{2}}$ & & & & & & & &  \\
        \hline \hline
        $B^- \to \bar{D}^0 K^-$    & & $1$& & $1$ & & & $1$ & & & & $1$ \\
\hline
$B^- \to D^- \bar{K}^0$    & & & & $-1$ & & &  & & & & $-1$   \\
\hline
$B^- \to D_s^- \pi^0$    & $\frac{1}{\sqrt{2}}$ & & & & $\frac{1}{\sqrt{2}}$\\
       \hline
       $ \bar{B}^0_s \to  D^-_s K^+$ & $1$ & & $1$& & $1$ & $1$ & & & $1$ & $1$\\
 \hline
 $\bar{B}^0_s \to D^- \pi^+$  & & & $1$& & & & & & $1$\\
 \hline
 $\bar{B}^0_s \to \bar{D}^0 \pi^0$  & & & $-\frac{1}{\sqrt{2}}$ & & & & & & $-\frac{1}{\sqrt{2}}$ \\
        \hline
        $ \bar{B}^0 \to  \bar{D}^0 \bar{K}^0$ & & $1$ & & & & & $1$ \\
 \hline
 $\bar{B}^0 \to D^-_s \pi^+$  & $1$ & & & & $1$\\
       \hline
    \end{tabular}
    \caption{Decomposition of the ${\bar B}\to {\bar D}P$ decay amplitudes in terms of effective diagrams. The first eight amplitudes are $\Delta S =0$ ($\propto V_{ub}^{}V^*_{cd}$), while the last eight amplitudes are $\Delta S =1$ ($\propto V_{ub}^{}V^*_{cs}$). \\}
    \label{tab:Diag_contrib_DC2}
\end{table}

\section{Observables}
\label{Sec:observables}

In order to perform fits of ${\bar B}\to D P$ and/or ${\bar B}\to {\bar D} P$ decays, we must gather up all the observables measured in these decays. We must also express these observables in terms of our unknown parameters, the magnitudes and relative phases of the diagrams.

Consider first the decay ${\bar B}\to f$, where $f$ is any $D P$ or ${\bar D} P$ final state. There are two observables associated with this decay. First, there is the branching ratio, defined as $\mathcal{B}({\bar B}\to f) = |A_f|^2 \Phi $, where $A_f$ is the amplitude for ${\bar B}\to f$ and $\Phi$ is the phase-space  factor. For $f = D P$, 
\beq \
\label{eq:Br}
\Phi = \frac{\sqrt{(m_B^2-(m_D-m_P)^2)(m_B^2-(m_D+m_P)^2)}}{16\pi m_B^3 \Gamma_B} ~,
\eeq
where $\Gamma_B$ is the decay width of the $B$. Second, there is the direct CP asymmetry, $\mathcal{A_{CP}}$, defined as
\beq
\mathcal{A_{CP}} \equiv 
\frac{\Gamma({\bar B}\to f) - \Gamma(B\to {\bar f})}
{\Gamma({\bar B}\to f) + \Gamma(B\to {\bar f})} ~,
\eeq
where ${\bar f}$ is the CP conjugate of $f$.

A direct CP asymmetry can be generated only if the decay ${\bar B}\to f$ involves at least two contributing amplitudes with nonzero relative weak and strong phases. However, all ${\bar B}\to DP$ decays have a single amplitude, proportional to $V^{}_{cb}V_{ud}^*$ ($\Delta S =0$) or $V^{}_{cb}V_{us}^*$ ($\Delta S =1$). This also holds for ${\bar B}\to {\bar D} P$ decays. Thus, these decays have no direct CP asymmetries.

\begin{table}[h!]
    \centering
    \begin{tabular}{|c|c|c|}
    \multicolumn{3}{c}{${\bar B}\to DP$: $\Delta S=0$ [$\propto V^{}_{cb}V_{ud}^* = \mathcal{O}(\lambda^2)$]}\\
    \hline \hline
        Decay mode & Branching Ratio  & Diagrams \\
        \hline
        \hline
         $B^- \to D^0 \pi^-$   &   ($4.61 \pm 0.10) \times 10^{-3}$    &      ${ T},{ C}$  \\
         \hline     
          $\bar{B}^0 \to D^+ \pi^-$   &        ($2.51 \pm 0.08) \times 10^{-3}$  & ${ T},{ E}$    \\
         \hline
          $\bar{B}^0 \to D^0 \pi^0$   &        ($2.67 \pm 0.09) \times 10^{-4}$ & ${ C},{ E}$   \\
         \hline
          $\bar{B}^0 \to D^+_s K^-$   &        ($2.7 \pm 0.5) \times 10^{-5}$ &  ${ E}$   \\
         \hline
          $\bar{B}^0_s \to D^0 K^0$   &       ($4.3 \pm 0.9) \times 10^{-4}$  & ${ C}$   \\
         \hline
         $\bar{B}^0_s \to D^+_s\pi^-$   &       ($2.98 \pm 0.14) \times 10^{-3}$  &  ${ T}$    \\
         \hline
         \end{tabular}
         ~~
            \begin{tabular}{|c|c|c|}
             \multicolumn{3}{c}{${\bar B}\to DP$: $\Delta S=1$  [$\propto V_{cb}^{}V_{us}^* = \mathcal{O}(\lambda^3)]$}\\
    \hline \hline
        Decay mode & Branching Ratio  & Diagrams \\
        \hline
        \hline
         $B^- \to D^0 K^-$   &       ($3.64 \pm 0.15) \times 10^{-4}$ & ${ T},{ C}$   \\
         \hline
          $\bar{B}^0 \to D^+ K^-$   &       ($2.05 \pm 0.08) \times 10^{-4}$ & ${ T}$    \\
         \hline
          $\bar{B}^0 \to D^0 \bar{K}^0$   &    $(5.5 \pm 0.4)\times 10^{-5}$     & ${ { C}}$  \\
         \hline
          $\bar{B}^0_s \to D^+ \pi^-$   & ---  & ${ E}$   \\
         \hline
          $\bar{B}^0_s \to D^0 \pi^0$   &  ---  & ${ E}$   \\
         \hline
         $\bar{B}^0_s \to D^+_s K^-$   &        ($1.94 \pm 0.21) \times 10^{-4}$ & ${ T},{ E}$  \\ 
         \hline
         \end{tabular}
    \caption{Branching ratios for $\bar{B}\to DP$ decays. The measured values of the branching ratios are taken from Ref.~\cite{ParticleDataGroup:2024cfk}, except for ${\mathcal{B}}({\bar B^0_s}\to  D_s^+K^-)$, which is taken from Ref.~\cite{Fleischer:2021cct}. `---' indicates that the branching ratio has not been measured.}
    \label{tab:Br_data_DC1}
\end{table}

\begin{table}[h!]
    \centering
    \begin{tabular}{|c|c|c|}
    \multicolumn{3}{c}{${\bar B}\to {\bar D}P: \Delta S=0$  [$\propto V_{ub}^{}V_{cd}^* = \mathcal{O}(\lambda^4)]$}\\
    \hline \hline
        Decay mode & Branching Ratio  &  Diagrams \\
         \hline
         $B^- \to \bar{D}^0 \pi^-$   & ---  &  ${{C}}',{{A}}'$ \\
         \hline
         $B^- \to D^- \pi^0$   &   ---   & ${{T}}',{{A}}'$   \\
         \hline
         $B^- \to D_s^- K^0$   &   $< 3.5 \times 10^{-6}$   & $ {{A}}'$  \\
         \hline
         $ \bar{B}^0 \to D^- \pi^+$   &   $(7.3 \pm 1.2)\times 10^{-7}$   &${{T}}',{{E}}'$  \\
         \hline
         $ \bar{B}^0 \to \bar{D}^0 \pi^0$   &   ---  & ${{C}}',{{E}}'$    \\
         \hline
         $ \bar{B}^0 \to D^-_s K^+$   &   ---   & ${{E}}'$    \\
         \hline
         $ \bar{B}^0_s \to D^- K^+$   &   ---   &    ${{T}}'$    \\
         \hline
         $ \bar{B}^0_s \to \bar{D}^0 K^0$   &   ---   &  ${{C}}'$   \\
         \hline
    \end{tabular}
    ~~
        \begin{tabular}{|c|c|c|}
    \multicolumn{3}{c}{${\bar B}\to {\bar D}P: \Delta S=1$  [$\propto V_{ub}^{}V_{cs}^* = \mathcal{O}(\lambda^3)]$}\\
    \hline \hline
        Decay mode & Branching Ratio  &  Diagrams \\
         \hline
         $B^- \to \bar{D}^0 K^-$   &   $(3.60 \pm 0.24)\times 10^{-6}$   &   ${{C}}',{{A}}'$ \\
         \hline
         $B^- \to D^- \bar{K}^0$   &   $< 2\times 10^{-6}$   &   ${{A}}'$   \\
         \hline
         $B^- \to D_s^- \pi^0$   &   $(1.6 \pm 0.5)\times 10^{-5}$    & ${{T}}'$  \\
         \hline
         $\bar{B}^0 \to \bar{D}^0 \bar{K}^0 $   &   ---   &  ${{C}}'$  \\
         \hline
         $\bar{B}^0 \to D_s^- \pi^+$   &   $(2.03 \pm 0.18)\times 10^{-5}$    & ${{T}}'$  \\
         \hline
         $\bar{B}^0_s \to D_s^- K^+$   &   $(2.6\pm 1.2)\times 10^{-5}$       & ${{T}}',{{E}}'$ \\
         \hline
         $\bar{B}^0_s \to D^- \pi^+$   &   ---   &     ${{E}}'$   \\
         \hline
         $\bar{B}^0_s \to \bar{D}^0 \pi^0$   &   ---   &  ${{E}}'$    \\
         \hline
    \end{tabular}
\caption{Branching ratios for ${\bar B}\to {\bar D}P$ decays. The measured values of the branching ratios are taken from Ref.~\cite{ParticleDataGroup:2024cfk}, except for ${\mathcal{B}}({\bar B}_s^0 \to D_s^- K^+)$, which is taken from Ref.~\cite{Fleischer:2021cct}. `---' indicates that the branching ratio has not been measured.}
    \label{tab:DC2_Br}
\end{table}

The measured branching ratios for $\bar{B}\to DP$ and ${\bar B}\to {\bar D}P$ decays are shown in Tables \ref{tab:Br_data_DC1} and \ref{tab:DC2_Br}, respectively. In these Tables, we estimate the sizes of the various CKM factors by writing them as powers of the Wolfenstein parameter $\lambda \simeq 0.2$. We see that the branching ratios for those decays whose amplitudes are proportional to $V_{cb}^{}V_{ud}^*$ [$= \mathcal{O}(\lambda^2)$] have all been measured, while for the decays whose amplitudes are proportional to $V_{ub}^{}V_{cd}^*$ [$= \mathcal{O}(\lambda^4)$], almost no branching ratios have been measured. In these Tables, we also list the SU(3)$_F$-invariant diagrams that contribute to each decay. In Ref.~\cite{Gronau:1994rj}, in which these diagrams were introduced, some naive estimates of ratios of diagrams were made: $|C^{(\prime)}/T^{(\prime)}| \simeq 1/3$ and 
$|E^{(\prime)}/T^{(\prime)}|, ~|A^{(\prime)}/T^{(\prime)}|  \ll 1$. In these Tables, we see that these estimates are borne out. For a given CKM factor, the branching ratios of decays dominated by $C^{(\prime)}$ are about an order of magnitude smaller than those dominated by $T^{(\prime)}$. And the branching ratios of decays dominated by $E^{(\prime)}$ or $A^{(\prime)}$ are about two orders of magnitude smaller than those dominated by $T^{(\prime)}$.

In fact, the statement that ``these decays have no direct CP asymmetries'' is not entirely true. From the above Tables, we see that $B^-$ decays to both $\bar{D}^0 K^-$ and $D^0 K^-$. The first amplitude is proportional to $V_{ub}^{}V_{cs}^*$, the second to $V_{cb}^{}V_{us}^*$. If both $\bar{D}^0$ and $D^0$ decay to the same final state $f'$, the decay is effectively $B^- \to K^- f'$, and there are two paths to get there, via the intermediate $\bar{D}^0$ or $D^0$. These two amplitudes have nonzero relative weak and strong phases, so that their interference will generate a direct CP asymmetry. 

There are two possibilities for $f'$. The first is that $f'$ is a CP eigenstate, either CP even or CP odd. In this case, the decaying neutral $D$ meson is also a CP-even or CP-odd CP eigenstate:
\beq 
\label{eq: Dcp}
D^{CP}_+ = \frac{1}{\sqrt{2}} (D^0 - {\bar D}^0) ~~,~~~~
D^{CP}_- = \frac{1}{\sqrt{2}} (D^0 + {\bar D}^0) ~,
\eeq
and there will be a direct CP asymmetry in $B^\pm \to D^{CP}_{+,-} K^\pm $. This also holds for $B^\pm \to D^{CP}_{+,-} \pi^\pm $. These decay modes are often used to extract the CP phase $\gamma$ \cite{Gronau:1990ra, Gronau:1991dp}. The second possibility is that $f' = \pi^\pm K^\mp$. Here there is a direct CP asymmetry in $B^\pm \to (\pi^\pm K^\mp) P^\pm $, $P = K, \pi$. These decays are also used to measure $\gamma$ \cite{Atwood:1996ci}.

Direct CP asymmetries are largest when the interfering amplitudes are of comparable size; the more the two amplitudes differ in size, the smaller is the asymmetry. The $B^- \to D^0 \pi^-$ amplitude is of $\mathcal{O}(\lambda^2)$ ($V_{cb}^{}V_{ud}^*$), while $B^- \to {\bar D}^0 \pi^- $ is of $\mathcal{O}(\lambda^4)$ ($V_{ub}^{}V_{cd}^*$). We therefore expect all direct CP asymmetries in $B^- \to D \pi^-$ decays to be small. Similarly, in $B^- \to (\pi^\pm K^\mp) K^- $, the $D^0 \to \pi^+ K^-$ amplitude is of $\mathcal{O}(1)$ ($V_{cs}^{}V_{ud}^*$), while ${\bar D}^0 \to \pi^+ K^-$ is of $\mathcal{O}(\lambda^2)$ ($V_{cd}^{}V_{us}^*$). The direct CP asymmetry in $B^- \to (\pi^\pm K^\mp) K^- $ is therefore also expected to be small. On the other hand, in $B^- \to D^{CP}_{+,-} K^- $, the CKM factors of all the interfering amplitudes are of the same order in $\lambda$, leading to a sizeable direct CP asymmetry. The measured values of these asymmetries are shown in Table \ref{tab:observables_directCPV}. We see that all of these expectations are borne out.

\begin{table}[h]
\begin{center}
    \begin{tabular}{|c|c|}
    \hline \hline
        Decay mode &    $\mathcal{A_{CP}}$ \\
        \hline
         $B^- \to (\pi^\pm K^\mp) \pi^- $   &   $-0.003 \pm 0.005$ \\
         $B^- \to D^{CP}_+ \pi^- $   &  $-0.008\pm0.003$ \\
         $B^- \to D^{CP}_- \pi^- $   &  $0.017\pm 0.026$ \\
        \hline
        $B^- \to (\pi^\pm K^\mp) K^-$ & $-0.017 \pm 0.005$ \\
        $B^- \to D^{CP}_+ K^- $ & $0.136\pm0.009$ \\
        $B^- \to D^{CP}_- K^- $ & $-0.14\pm 0.05$ \\ 
         \hline
    \end{tabular}
\end{center}
\caption{Direct CP asymmetries in $B^- \to D P^-$ decays, $P=\pi, K$. All data is taken from Ref.~\cite{ParticleDataGroup:2024cfk}.\label{tab:observables_directCPV}}
\end{table}

Finally, we turn to the decays of neutral ${\bar B}$ mesons. Because of $B^0$-${\bar B}^0$ mixing, the state that decays is a time-dependent mixture of $B^0$ and ${\bar B}^0$. If the final state $f$ is accessible to both ${\bar B}^0$ and $B^0$, in the time-dependent decay rate, the interference between ${\bar B}^0 \to f$ and ${\bar B}^0 \to B^0 \to f$ can lead to CP-violating effects. 

As was the case for neutral $D$ decays, $f$ can be a CP eigenstate or a non-CP eigenstate. For $B^0$/${\bar B}^0$ decays, the non-CP eigenstates include $ D^\mp \pi^\pm $ \cite{Dunietz:1987bv, Fleischer:2003yb}, $D^0 \pi^0$ \& $ {\bar D}^0 \pi^0 $ \cite{Fleischer:2003ai}, $D^0 K_S $ \& $ {\bar D}^0 K_S $ \cite{Gronau:1990ra, Kayser:1999bu, Atwood:2002vw, Fleischer:2003aj}; for $B^0_s$/${\bar B}^0_s$ decays, they are $D^\mp_s K^\pm $ \cite{Aleksan:1991nh, DeBruyn:2012jp, Fleischer:2021cct}, $D^0 K_S$ \& ${\bar D}^0 K_S $ \cite{Fleischer:2003aj}. The CP eigenstates for both $B^0$/${\bar B}^0$ and $B^0_s$/${\bar B}^0_s$ decays are $ D^{CP} \pi^0 $ \cite{Fleischer:2003ai} and $D^{CP} K_S $ \cite{Gronau:1990ra, Fleischer:2003ai}. 

Including both $B^0$-${\bar B}^0$ mixing and the fact that both $B^0$ and ${\bar B}^0$ can decay to the final state $f$, one can construct the time-dependent CP asymmetry. For the case where $f$ is a non-CP eigenstate, we define two parameters: 
\beq
\lambda_f \equiv \frac{q}{p} \, \frac{\bar A_f}{A_f} ~~,~~~~ 
\lambda_{\bar f} \equiv \frac{q}{p} \, 
\frac{\bar A_{\bar f}}{A_{\bar f}} ~.
\eeq
Here, the factor $q/p$ comes from $B^0$-$\bar{B}^0$ mixing. For $B^0$ ($B_s^0$) mesons, it is equal to $\exp(-2\beta)$ [$\exp(-2\beta_s)$], so $|q/p| = 1$. In addition,  
\beq
A_f \equiv A(B^0 \to f) ~~,~~~~ 
{\bar A}_f \equiv A({\bar B}^0 \to f) ~~,~~~~ 
A_{\bar f} \equiv A(B^0 \to {\bar f}) ~~,~~~~ 
{\bar A}_{\bar f} \equiv A({\bar B}^0 \to {\bar f}) ~.
\eeq
The time-dependent CP asymmetry is then given by (e.g., see Eqs.~(115) and (116) in Ref.~\cite{HFLAV:2022esi})
\begin{eqnarray}
\label{eq:A_CP}
 \mathcal{A}_f(t) & = \frac{\Gamma_{\bar B \to f}(t) - \Gamma_{ B \to f}(t)}{\Gamma_{\bar B \to f}(t) + \Gamma_{ B \to f}(t)} = S_f \sin(\Delta mt) - C_f\cos(\Delta m t) ~, \nn\\
 \mathcal{A}_{\bar f}(t) & = \frac{\Gamma_{\bar B \to {\bar f}}(t) - \Gamma_{ B \to {\bar f}}(t)}{\Gamma_{\bar B \to {\bar f}}(t) + \Gamma_{ B \to {\bar f}}(t)} = S_{\bar f} \sin(\Delta mt) - C_{\bar f}\cos(\Delta m t) ~.
\end{eqnarray}
The coefficients $C_{f(\bar f)}$ and $S_{f(\bar f)}$ are what is measured experimentally; they are, respectively, the direct and indirect CP asymmetries. Their expressions are given by
\begin{eqnarray}
    C_{f(\bar f)} & = &\frac{1 - |\lambda_{f(\bar f)}|^2}{1 + |\lambda_{f(\bar f)}|^2} 
    = \frac{|A_{f({\bar f})}|^2 - |{\bar A}_{f({\bar f})}|^2}
    {|A_{f({\bar f})}|^2 + |{\bar A}_{f({\bar f})}|^2} ~, \nn\\
    S_{f(\bar f)} & = & \frac{2\,{\rm Im}(\lambda_{f(\bar f)})}{1+|\lambda_{f(\bar f)}|^2} 
    = \frac{2}{|A_{f({\bar f})}|^2 + |{\bar A}_{f({\bar f})}|^2} \, {\rm Im} \left( \frac{q}{p} \, {\bar A}_{f({\bar f})} A_{f({\bar f})}^* \right) ~.
    \label{DirectCPAdef}
\end{eqnarray}
If $f$ is a CP eigenstate, the derivation of the direct and indirect CP asymmetries is similar, except that $f = {\bar f}$.

Only a few measurements of the time-dependent CP asymmetries have been made. The results are shown in Table \ref{tab:observables_indirectCPV}. A few words of explanation are warranted. In the decays $\bar{B}^0 \to \pi^\pm D^\mp$, the two amplitudes $A_f$ and ${\bar A}_f$ are very different in size -- one is $\mathcal{O}(\lambda^2)$ and the other is $\mathcal{O}(\lambda^4)$. As a result, the direct CP asymmetry $C_f$ [Eq.~(\ref{DirectCPAdef})] is $\simeq \pm 1$. Similarly, the indirect CP asymmetry $S_f$ is very small. By contrast, in $\bar{B}^0_s \to K^\pm D^\mp_s$, $A_f$ and ${\bar A}_f$ are similar in size, leading to sizeable $C_f$ and $S_f$.

\begin{table}[h]
\begin{center}
    \begin{tabular}{|c||c|c|}
    \hline \hline
        Decay mode & $C$ & $S$  \\
        \hline
    $\bar{B}^0 \to D^+ \pi^-$   & $-1$  & $0.038 \pm 0.021$ \\ $\bar{B}^0 \to D^- \pi^+$   & $1$   & $0.058 \pm 0.023$ \\
$\bar{B}^0_s \to D^+_s K^-$ & $-0.73 \pm 0.15$ & $ -0.49 \pm 0.21$ \\ 
$\bar{B}^0_s \to D_s^- K^+$ & $0.73 \pm 0.15$ & $-0.52 \pm 0.21$    \\  
         \hline
    \end{tabular}

\end{center}
\caption{Direct and indirect CP-asymmetries in neutral $B$-meson decays. All data is taken from Ref.~\cite{ParticleDataGroup:2024cfk}. \label{tab:observables_indirectCPV}}
\end{table}

\section{Fits to the Data}
\label{sec:fits}

In Sec.~\ref{Sec: amplitudes}, we expressed all ${\bar B}\to D P$ and ${\bar B}\to {\bar D} P$ decay amplitudes as functions of SU(3)$_F$-invariant and SU(3)$_F$-breaking diagrams. We can use these expressions to fit the data, using the magnitudes and relative phases of the diagrams as the unknown fit parameters. In this section, we present the results of the following three types of fits. First, we consider ${\bar B}\to D P$ decays alone, essentially updating the analysis in Ref.~{\cite{Davies:2024vmv}. We refer to these as Fit 1. Next, we examine ${\bar B}\to {\bar D} P$ decays alone -- these are referred to as Fit 2. Finally, we combine both ${\bar B}\to D P$ and ${\bar B}\to {\bar D} P$ decays in Fit 3. Using the \emph{Minuit} package \cite{James:1975dr} we find the minimum value of the $\chi^2$ function, $\chi^2_{\rm min}$, for each fit. We then utilize this $\chi^2_{\rm min}$ value as a statistic for hypothesis testing.

Our goal is to test the following hypotheses using the $\chi^2_{\rm min}$ obtained in each fit: 
\begin{enumerate}[label=(\alph*)]

\item \emph{that SU(3)$_F$ breaking is not required to explain the data}, 

\item \emph{that the data can be explained by including up to 30\% SU(3)$_F$ breaking}, 

\item \emph{that at least 60\% SU(3)$_F$ breaking is needed to explain the data.} 

\end{enumerate}
Along the way, we sometimes perform additional fits not necessarily aligned with one of these hypotheses, but rather to better understand the results of a fit. Inspired by the results of Ref.~\cite{Davies:2024vmv}, we treat hypothesis (b) as our \emph{null hypothesis}. In each type of fit, hypothesis (a) applies the largest symmetry group and therefore utilizes the fewest unknown fit parameters. The other hypotheses within each fit then require additional fit parameters (the SU(3)$_F$-breaking diagrams). 

In Fits 1 and 3, there are more observables than parameters for hypothesis (a), so that the number of degrees of freedom in these fits is greater than zero. However, this is not the case for Fit 2, where there is not enough data to perform a conventional fit\footnote{In a scenario in which all the observables depend on linear combinations of the fit parameters, and in which there are more parameters than observables, one expects infinitely many solutions confined to an ``$n$''-dimensional plane, where $n$ refers to the difference between the number of parameters and observables. In our case, however, the relationship between parameters and observables is highly nonlinear, often including inverse trigonometric functions. As such, it is possible that, even in scenarios involving overfitting where the number of parameters exceeds the number of observables, an exact solution does not exist.} with hypothesis (a). In other words, in Fit 2, even hypothesis (a) requires an overfitting of the data. Invoking hypothesis (b) automatically leads to additional parameters but not additional data. However, under circumstances discussed in this section, tensions with SU(3)$_F$ limit arise even in these situations. This tension is somewhat alleviated by the addition of SU(3)$_F$ breaking. 

\begin{table}[!htbp]
\begin{center}
\begin{tabular}{c c | c c c | c} \hline \hline
\multicolumn{2}{c|}{Fit} & \multicolumn{3}{c|}{Number of} & \\
Type & Hypothesis & Observables & Parameters & d.o.f. & $\chi^2_{\rm min}$ \\ \hline \hline
1 & (a) & 10 & 5 & 5 & 105.0 \\
1 & (b) & 10 & 15 & -- & 0.0 \\ \hline \hline
2 & (a) & 5 & 7 & -- & 2.0 \\ 2 & (b) & 5 & 19 & -- & 1.0 \\ \hline\hline
3 & (a) & 25 & 13 & 12 & 153.4 \\
3 & (b) & 25 & 35 & -- & 17.0 \\
3 & (c) & 25 & 35 & -- & 2.8 \\ \hline \hline
\end{tabular}
\end{center}
\caption{Summary of various fits detailed in this section. Here d.o.f. stands for degrees of freedom. In fit scenarios involving overfitting, i.e., when the number of observables is less than the number of parameters, the d.o.f. column lists `--'.} \label{tab:fitsum}
\end{table}

In Table \ref{tab:fitsum}, we summarize the various fits corresponding to the different hypotheses used, the numbers of observables, parameters, and degrees of freedom (d.o.f.), and the $\chi^2_{\rm min}$ values. In Fit 1, we find that hypothesis (b), our null hypothesis, agrees with the data. This renders hypothesis (c) unnecessary to consider for Fit 1. This is also somewhat true for Fit 2, where we show that there is tension in the data that cannot be removed by adding SU(3)$_F$ breaking. In the rest of this section, we delve into each type of fit in greater detail.

\subsection{Fits 1: \boldmath ${\bar B}\to D P$ Decays}
\label{sec:fits1}

There are 12 ${\bar B}\to D P$ decays (see the left panel of Table \ref{tab:List_decays_B_DP}).  In the SU(3)$_F$ limit, their amplitudes are expressed in terms of three diagrams, $T$, $C$ and $E$. When one adds SU(3)$_F$-breaking diagrams, the amplitudes are written in terms of eight effective diagrams (see Table \ref{tab:Diag_contrib_DC1}). Three of these -- $\tilde{T}$, $\tilde{C}$ and $\tilde{E}$ -- are the SU(3)$_F$-invariant diagrams with an SU(3)$_F$-breaking correction, while the other five are simply combinations of SU(3)$_F$-breaking diagrams. Since we are focusing only on ${\bar B}\to D P$ decays, the only observables are the branching ratios. As shown in Table \ref{tab:Br_data_DC1}, 10 of these branching ratios have been measured. In Sec.~\ref{Sec:observables}, all observables are defined in terms of amplitudes. This allows us to express the branching ratios for all decays in terms of diagrams and perform fits.

\subsubsection{Hypothesis (a): exact SU(3)$_F$}
\label{sec:fits1a}

To test hypothesis (a), we perform a fit to the ${\bar B}\to D P$ decay observables expressed in terms of the three diagrams, $T$, $C$ and $E$. This corresponds to five unknown parameters -- three magnitudes and two relative strong phases. Ten observables (branching ratios) have been measured (Table \ref{tab:Br_data_DC1}), so the number of d.o.f.~is 5. \emph{Minuit} is used to minimize the $\chi^2$ function with respect to the five unknown fit parameters. We find $\chi^2_{\rm min}/{\rm d.o.f.} = 105/5$, with a corresponding $p$-value $\simeq 0$, which excludes hypotheses (a). The best-fit parameter values are shown in Table \ref{tab: B_DP_SU3_fit}.

\begin{table}[h!]
    \centering
    \begin{tabular}{c|c}
    \hline
    Parameter & Best-fit value \\
    \hline
    $|T|$ & $1.91 \pm 0.02$  \\
    $|C|$ & $0.77 \pm 0.04$  \\
    $|E|$ & $0.20 \pm 0.01$ \\
    \hline
    $\delta_C$ & $(-75 \pm 6)^\circ$  \\
    $\delta_E$ & $(-160 \pm 10)^\circ$ \\
    \hline     
    \end{tabular}
    \caption{Best-fit values of the parameters for the fit of ${\bar B}\to D P$ decays assuming exact SU(3)$_F$ symmetry. The magnitudes of the diagrams are given in units of $10^{-5}$ GeV. The strong phase $\delta_T$ is taken to be zero.}
    \label{tab: B_DP_SU3_fit}
\end{table}

The results shown in Table \ref{tab: B_DP_SU3_fit} above clearly demonstrate a severe departure from hypothesis (a). This was already noticed in Ref.~\cite{Davies:2024vmv}. Here we go one step further and investigate the reason behind this poor fit by analyzing the individual $\Delta S=0$ and $\Delta S=1$ ${\bar B}\to D P$ decays. Table \ref{tab:Br_data_DC1} lists six $\Delta S=0$ and six $\Delta S=1$ decays. Under hypothesis (a), all twelve decay amplitudes depend on the same five parameters. The branching ratios of all six $\Delta S=0$ decays have been measured. However, among the six $\Delta S=1$ decays, only four branching ratios have been measured. In each case, we perform a separate $\chi^2$ minimization fit, and here is what we find. 

The $\Delta S=0$ fit has one more observable than parameters. We find an excellent fit with $\chi^2_{\rm min}/{\rm d.o.f.} = 0.2/1$, corresponding to a $p$-value of 0.89. On the other hand, the $\Delta S=1$ fit has one more parameter than observables. This leads to an overfitting scenario for which one might expect multiple discretely ambiguous solutions. However, we find that only one set of parameter values yields an exact solution ($\chi^2 = 0$). For both sets of decays, the preferred values of the fit parameters are shown in Table \ref{tab: DeltaS01_fits}. We see that the largest diagram in the $\Delta S=0$ ($\Delta S=1$) fit is $T_0$ ($T_1$), and the error is quite small. The central value of $|T_1|$ is many standard deviations away from the central value of $|T_0|$, so forcing them to be equal, as is done in Fit 1 under hypothesis (a), will result in a very poor fit.
\begin{table}[h!]
    \centering
    \begin{tabular}{c|c}
    \multicolumn{2}{c}{$\Delta S = 0$ ${\bar B}\to D P$ decays}\\
    \hline
    Parameter & Best-fit value \\
    \hline
       $|T_0|$  & $1.84 \pm 0.03$ \\
        $|C_0|$ & $0.70 \pm 0.07$\\
        $|E_0|$ & $0.18 \pm 0.02$ \\
        \hline
        $\delta_{C_0}$ & $(-71 \pm 6)^\circ$ \\
        $\delta_{E_0}$ & $(-200 \pm 200)^\circ$\\
      \hline     
    \end{tabular}
    ~~~~~
        \begin{tabular}{c|c}
    \multicolumn{2}{c}{$\Delta S = 1$ ${\bar B}\to D P$ decays}\\
    \hline
    Parameter & Best-fit value \\
    \hline
        $|T_1|$  & $2.10 \pm 0.04$ \\
        $|C_1|$ & $1.1 \pm 0.1$\\
        $|E_1|$ & $0.1 \pm 0.7$ \\
        \hline
        $\delta_{C_1}$ & $(-67 \pm 6)^\circ$ \\
        $\delta_{E_1}$ & $(-100 \pm 300)^\circ$\\
    \hline     
    \end{tabular}
    \caption{Best-fit values of the parameters for the fits of $\Delta S=0$ and $\Delta S=1$ ${\bar B}\to D P$ decays, respectively on the left and right. The magnitudes of the diagrams are given in units of $10^{-5}$ GeV. The strong phases $\delta_{T_0}$ and $\delta_{T_1}$ are taken to be zero.}
    \label{tab: DeltaS01_fits}
\end{table}

There is another way of looking at these results. Consider the following comparisons between the sizes of the $\Delta S=1$ and $\Delta S=0$ diagrams: 
\beq |T_1/T_0| -1 = 0.15 \pm 0.03 ~,~~ |C_1/C_0|-1 = 0.36 \pm 0.13 ~,~~ |E_1/E_0| -1 = -0.2 \pm 3.3 ~.
\eeq 
These provide a rough idea of the size of SU(3)$_F$ breaking, suggesting that roughly 15\%-36\% SU(3)$_F$-breaking contributions to these diagrams may lead to far better fits. In the following subsection, we investigate this possibility by using hypothesis (b).

\subsubsection{Hypothesis (b): up to 30\% SU(3)$_F$ breaking}
\label{sec:fits1b}

A clear rejection of hypothesis (a) in Fit 1 paves the way for the consideration of hypothesis (b), ``that the data can be explained by including up to 30\% SU(3)$_F$ breaking.'' To implement hypothesis (b) in Fit 1, we need to expand our parameter space to include SU(3)$_F$-breaking diagrams. In section \ref{sec:formalism}, it was shown that, once first-order SU(3)$_F$ breaking ($\propto (m_s - m_d)/\Lambda_{\rm QCD}$) is included, there are eight independent combinations of diagrams. Of these, three diagram combinations -- $\tilde{T}$, $\tilde{C}$, and $\tilde{E}$ -- replace $T$, $C$, and $E$. Another five diagrams -- $\tilde{T}_{1,2}, \tilde{C}_{1,2}$, and $\tilde{E}_1$ -- arise from SU(3)$_F$ breaking. There are therefore a total of 15 parameters -- eight magnitudes and seven relative strong phases.

To perform a fit consistent with hypothesis (b), the null hypothesis, we need to place limits on the sizes of the breaking diagrams relative to their corresponding unbroken contributions. This is done by defining a new set of breaking parameters consisting of magnitude ratios, $r_{X_i} = |\tilde{X}_i/\tilde{X}|$, and phase differences, $\delta_{X_i} = {\rm arg}(\tilde{X}_i) - {\rm arg}(\tilde{X})$. The 15 parameters used in our fit are then the magnitudes $|\tilde{T}|$, $|\tilde{C}|$, $|\tilde{E}|$, the ratios $r_{T_{1,2}}$, $r_{C_{1,2}}$, $r_{E_1}$, and the relative phases $\delta_{\tilde{C}}, \delta_{\tilde{E}}, \delta_{\tilde{T}_{1,2}}, \delta_{\tilde{C}_{1,2}}$, and  $\delta_{\tilde{E}_{1}}$. As before, we use \emph{Minuit} to find the minimum of the $\chi^2$ function. However, here we also place limits on all ratio parameters, $r_{X_i} < 30\%$. In Table \ref{tab:DC1_breaking_fit} we present the results of our fit. We find $\chi^2_{\rm min} = 0$, which is obviously a good fit. The conclusion from this fit is that the null hypothesis successfully explains the data.

\begin{table}[!htbp]
\begin{center}
    \begin{tabular}{c|c|c|c}
     \hline
        Magnitude  & Best-fit value & Phase & Best-fit value \\
        \hline
        $|\tilde{T}|$ & $1.8 \pm 0.1 $ & -  &  - \\
        $|\tilde{C}|$ & $0.9 \pm 0.1 $ & $\delta_{\tilde{C}}$  &  $(-74\pm 4)^\circ$ \\
        $|\tilde{E}|$ & $0.18 \pm 0.02 $ & $\delta_{\tilde{E}}$  &  $(-130\pm 40)^\circ$ \\
        $r_{T_1}$ & $0.2 \pm 0.2 $ & $\delta_{T_1}$  &  $(0\pm 300)^\circ$ \\
        $r_{T_2}$ & $0.0 \pm 0.3 $ & $\delta_{T_2}$  &  $(0\pm 300)^\circ$ \\
        $r_{C_1}$ & $0.2 \pm 0.3 $ & $\delta_{C_1}$  &  $(0\pm200)^\circ$ \\
        $r_{C_2}$ & $0.2 \pm 0.2 $ & $\delta_{C_2}$  &  $(-200\pm200)^\circ$ \\
        $r_{E_1}$ & $0.2 \pm 0.2 $ & $\delta_{E_1}$  &  $(-200\pm300)^\circ$ \\
        \hline
    \end{tabular}
\end{center}
\caption{Best-fit values of the 15 fit parameters for hypothesis (b). The phase of $\tilde{T}$ has been set to zero. The magnitudes of the diagrams -- $|\tilde{T}|$, $|\tilde{C}|$, and $|\tilde{E}|$ -- are given in units of $10^{-5}$ GeV. \label{tab:DC1_breaking_fit}}
\end{table}

Despite the acceptance of the null hypothesis, a key question remains. Is this agreement with the data simply a consequence of the expanded parameter space? After all, our fit has 15 parameters that describe only 10 observables, and as such one might expect that multiple sets of parameter values can explain the data. Even so, the results of the fit make it clear that, when the five SU(3)$_F$-breaking ratio parameters are floated with an upper bound of 30\%, their fit values settle at around 20\%, albeit with large error bars.

To explore this further, we implement several additional fits by tweaking hypothesis (b). Instead of a hard 30\% upper bound, we constrain the upper bound on all ratio parameters to be $r_{X_i} \leq R$, where we vary $R$ from 0 to 30\% ($R = 0$ and $R = 30\%$ represent hypothesis (a) and hypothesis (b), respectively). In Fig.~\ref{fig:chi2Rf1}, we show that the $\chi^2_{\rm min}$ decreases as $R$ increases from 0 to $\sim 20$\%. When $R = 0$, the fit resembles hypothesis (a) -- we find $\chi^2_{\rm min} \sim 105$. For $R > 20\%$, the fits resemble hypothesis (b) -- we find $\chi^2_{\rm min} \sim 0$. However, for $R \in (0,\sim 0.2)$, the fits find a nonzero $\chi^2$. This demonstrates that it is not sufficient to simply enlarge the parameter space to obtain a good fit; we need to allow $\gtrsim 20\%$ SU(3)$_F$ breaking in order to explain the data. 

Note that our results here are very similar to those obtained in Ref.~\cite{Davies:2024vmv}. However, a key difference between our formalisms is that we use one fewer parameter to test the null hypothesis. In our case, we have established the equivalence between diagrams and RMEs (see Appendix \ref{RME_Diag_Equiv}) which has led to the reabsorption of one SU(3)$_F$-breaking diagram.

\begin{figure}[!htbp]
\centering
\includegraphics[width=0.75\textwidth]{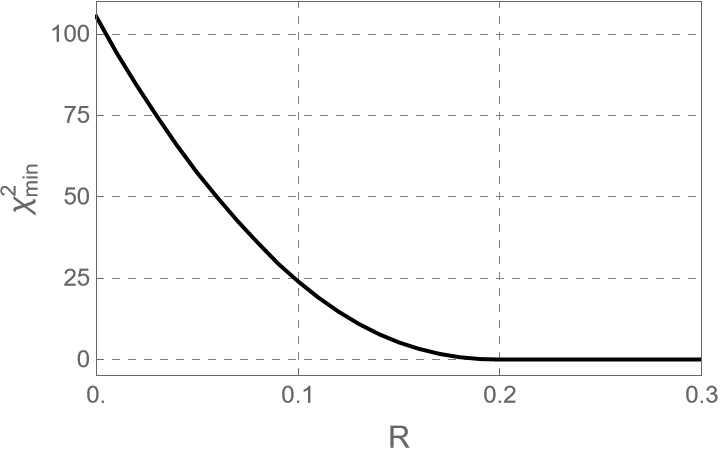}
\caption{$\chi^2_{\rm min}$ as a function of $R$ for fits to ${\bar B}\to DP$ data. Here $R$ is the upper limit set on all ratios $r_{X_i}$ in a given fit.
\label{fig:chi2Rf1}}
\end{figure}

In summary, in studying ${\bar B}\to DP$ decays alone, we have demonstrated that the exact SU(3)$_F$ scenario of hypothesis (a) is quite far from the data. We have also shown that the addition of $\gtrsim 20\%$ SU(3)$_F$ breaking is sufficient to explain the data. We therefore conclude that the null hypothesis cannot be rejected with data from ${\bar B}\to DP$ decays alone. 

\subsection{Fits 2: \boldmath ${\bar B}\to{\bar D}P$ Decays}
\label{sec:fits2}

There are 16 ${\bar B}\to{\bar D}P$ decays (see the right panel of Table \ref{tab:List_decays_B_DP}). In the SU(3)$_F$ limit, their decay amplitudes depend on four diagrams -- $T', C', E'$, and $A'$. When SU(3)$_F$-breaking diagrams are included, the number of effective diagrams grows to 11 (see Table \ref{tab:Diag_contrib_DC2}). Four of these -- $\tilde{T}', \tilde{C}', \tilde{E}', \tilde{A}'$ -- are the SU(3)$_F$-invariant diagrams with an SU(3)$_F$ breaking correction, while the remaining seven are combinations of SU(3)$_F$-breaking diagrams. When we focus only on ${\bar B}\to{\bar D}P$ decays, branching ratios are again the only observables. However, as shown in Table \ref{tab:DC2_Br}, only five of the 16 branching ratios have been measured, and two others have upper bounds. Using the expressions for the observables in terms of diagrams, we once again perform fits to test our hypotheses.

\subsubsection{Hypothesis (a): exact SU(3)$_F$}
\label{sec:fits2a}

Under the assumption of exact SU(3)$_F$ symmetry, the five measured ${\bar B}\to{\bar D}P$ branching fractions can be expressed in terms of four diagrammatic amplitudes. The $\chi^2$ function therefore depends on seven parameters -- four magnitudes and three relative strong phases. Using \emph{Minuit}, we minimize the $\chi^2$ function and find $\chi^2_{\rm min} = 2.0$. The best-fit values of the parameters are listed in Table \ref{tab:DC2_SU(3)_fit_diag}. Two additional branching ratios have upper bounds. Both of these channels are dominated by the annihilation amplitude $A'$. In our fits, we find that $|A'|$ is two orders of magnitude smaller than the dominant tree amplitude, $|T'|$. Our fit results thus comply well with these bounds.

\begin{table}[h!]
\begin{center}
\begin{tabular}{c|c} \hline
    Parameter  & Best-fit values \\
        \hline
    $|T'|$  & $1.41 \pm 0.04$ \\
    $|C'|$ & $0.6 \pm 0.1$\\
    $|E'|$ & $0.3 \pm 0.1$ \\
    $|A'|$ & $0.01 \pm 0.07$ \\ \hline
    $\delta_{C'}$ & $(100\pm200)^\circ$ \\
    $\delta_{E'}$ & $(-150\pm20)^\circ$\\
    $\delta_{A'}$ & $(100\pm300)^\circ$ \\ \hline
 \end{tabular}
 \end{center}
 \caption{Best-fit values of the parameters for the fit of ${\bar B}\to{\bar D}P$ decays assuming exact SU(3)$_F$ symmetry. The magnitudes of the diagrams are given in units of $10^{-5}$ GeV. The strong phase $\delta_{T'}$ is taken to be zero.}
 \label{tab:DC2_SU(3)_fit_diag}
\end{table}

Note that this fit involved five observables that depend on seven real parameters. This means that overfitting was necessary. In this case, one naturally expects to find $\chi^2_{\rm min} = 0$. However, we find a non-zero value of $\chi^2_{\rm min}$. It turns out that there are two sources of this nonzero minimum $\chi^2$. These are as follows.
\begin{itemize}
  
\item Under isospin symmetry, the amplitudes for $B^-\to D^-_s\pi^0$ and ${\bar B}^0\to D^-_s \pi^+$ are related. The difference in phase-space factors between these two decays, arising from the tiny mass differences between the charged and neutral mesons, is negligible. Based on this, one expects ${\cal B}({\bar B}^0\to D^-_s \pi^+) \simeq 2 {\cal B}(B^-\to D^-_s\pi^0)$. However, the measured central values are roughly 1$\sigma$ away from this expectation. This gives rise to a $\Delta\chi^2_{\rm min} \sim 1.0$. This contribution cannot be removed in our analysis as isospin breaking is necessary to resolve this discrepancy.

\item In the SU(3)$_F$ limit, the amplitudes for ${\bar B}^0\to D^-\pi^+$ and ${\bar B}^0_s\to D^-_sK^+$ are equal, as these two decays are related by U-spin symmetry. Apart from a small difference in phase-space factors (arising from the mass differences in the initial- and final-state mesons), the two branching ratios are expected to be similar in size. However, their measured values differ by two orders of magnitude. We find that this difference in branching ratios contributes $\Delta\chi^2_{\rm min} \sim 1.0$. We will see that this deviation can be removed by including SU(3)$_F$ breaking in hypothesis (b).

\end{itemize}

Now, in principle one can repeat the steps done in Sec.~\ref{sec:fits1a}, namely separately studying the $\Delta S = 0$ and $\Delta S = 1$ decays in the ${\bar B}\to{\bar D}P$ system. However, the problem is that not enough data is available. Of the eight $\Delta S = 0$ branching ratios, only one measurement and one upper bound are available. These depend on seven SU(3)$_F$ symmetry parameters, so a fit is not meaningful. The situation in $\Delta S = 1$ channels is slightly better, as there are four measurements and one upper bound that also depend on seven parameters. Still, our combined fit with hypothesis (a) already demonstrated the sources of the deviation. We therefore proceed to apply hypothesis (b) to this system.

\subsubsection{Hypothesis (b): up to 30\% SU(3)$_F$ breaking}
\label{sec:fits2b}

In the analysis of hypothesis (a) for Fit 2, we found that the $\chi^2_{\rm min}$ has a removable contribution from a pair of U-spin-related decays. To address this, here we use hypothesis (b). When first-order SU(3)$_F$-breaking contributions are included, there are 11 effective diagrams (see Table \ref{tab:Diag_contrib_DC2}) -- four unbroken-SU(3)$_F$ diagrams with SU(3)$_F$-breaking corrections, and seven combinations of SU(3)$_F$-breaking diagrams. Of these, $\tilde{C}'_2$ appears only in the amplitude of ${\bar B}_s^0\to K^0{\bar D}^0$. This branching ratio has not been measured, so there is no way to constrain $\tilde{C}'_2$. Dropping this diagram, there are now 10 effective diagrams leading to 19 parameters -- 10 magnitudes and nine relative strong phases. 

As before, we define the following SU(3)$_F$-breaking parameters: $r_{X'_i} = \tilde{X}'_i/\tilde{X}$ and $\delta_{X'_i} = {\rm arg}(\tilde{X}'_i) - {\rm arg}(\tilde{X}')$, where $X = T, C, E, A$, and $i = 1, 2$. We carry out a fit to minimize $\chi^2$. To implement hypothesis (b), we place an upper bound on the ratios: $r_{X'_i} \leq R = 30\%$. We find $\chi^2_{\rm min} = 0.98$; our results are presented in Table \ref{tab:DC2_SU(3)_breaking_diag}.

\begin{table}[h]
\begin{center}
 \begin{tabular}{c|c|c|c} \hline
    Parameter & Best-fit value & Parameter & Best-fit value \\ \hline
    $|\tilde{T}'|$ & $1.2\pm0.3$ & -- & -- \\
    $|\tilde{C}'|$ & $0.5\pm0.8$ & $\delta_{\tilde{C}'}$ & $(100\pm300)^\circ$ \\
    $|\tilde{E}'|$ & $0.1\pm0.7$ & $\delta_{\tilde{E}'}$ & $(100\pm200)^\circ$\\
    $|\tilde{A}'|$ & $0.0\pm0.1$ & $\delta_{\tilde{A}'}$ & $(0\pm300)^\circ$ \\
    $r_{T'_1}$ & $0.2\pm0.3$ & $\delta_{T'_1}$ & $(0\pm100)^\circ$ \\
    $r_{T'_2}$ & $0.2\pm0.3$ & $\delta_{T'_2}$ & $(0\pm300)^\circ$ \\
    $r_{C'_1}$ & $0.3\pm0.3$ & $\delta_{C'_1}$ & $(100\pm500)^\circ$ \\
    $r_{E'_1}$ & $0.3\pm0.2$ & $\delta_{E'_1}$ & $(-100\pm300)^\circ$ \\
    $r_{E'_2}$ & $0.1\pm0.2$ & $\delta_{E'_2}$ & $(-200\pm 300)^\circ$ \\
    $r_{A'_1}$ & $0.1\pm0.2$ & $\delta_{A'_1}$ & $(100\pm400)^\circ$ \\ \hline
\end{tabular}
\end{center}
\caption{Best-fit values of the 19 parameters for the fit of ${\bar B}\to{\bar D}P$ decays assuming hypothesis (b). The magnitudes of the diagrams are given in units of $10^{-5}$ GeV. The strong phase $\delta_{\tilde{T}'}$ is taken to be zero.
\label{tab:DC2_SU(3)_breaking_diag}}
\end{table}

\begin{figure}[!htbp]
\centering
\includegraphics[width=0.75\textwidth]{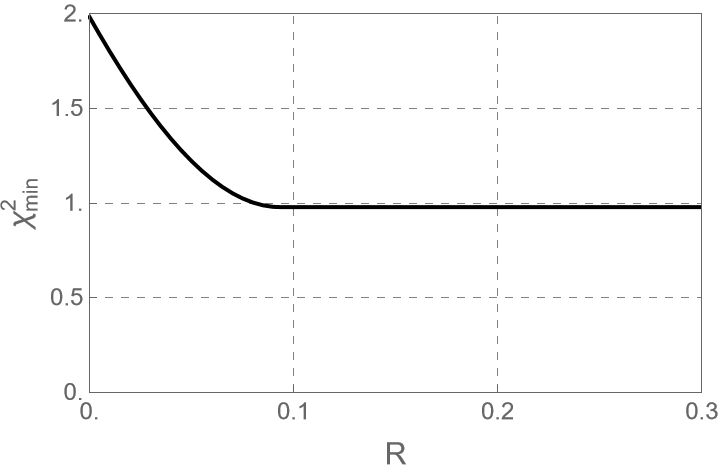}
\caption{$\chi^2_{\rm min}$ as a function of $R$ for fits to ${\bar B}\to {\bar D}P$ data. Here $R$ is the upper limit set on all ratios $r_{X'_i}$ in a given fit. There is a residual $\chi^2_{\rm min} = 1.0$ for $R \gtrsim 10\%$ that cannot be removed by adding further SU(3)$_F$ breaking. This residual $\chi^2_{\rm min}$ can be attributed to a deviation from isospin symmetry. 
\label{fig:chi2Rf2}}
\end{figure}

As was done in the analysis of ${\bar B}\to D P$ decays with hypothesis (b), we now vary $R$ from 0 to 30\%. In Fig.~\ref{fig:chi2Rf2}, we plot the value of $\chi^2_{\rm min}$ as a function of $R$. $R = 0$ represents the case with exact SU(3)$_F$ symmetry, and we find $\chi^2_{\rm min} = 2.0$. For $R \geq 10\%$, we find that $\chi^2_{\rm min} = 1.0$, the residual contribution coming from a difference between two isospin-related branching ratios. Thus, ignoring any isospin-breaking effects, one can conclude that 30\% SU(3)$_F$ breaking is sufficient to explain the available data in the ${\bar B}\to{\bar D}P$ system, i.e., the data confirms hypothesis (b). 

\subsection{Fits 3: \boldmath ${\bar B}\to D P$ and ${\bar B}\to {\bar D} P$ Combined}
\label{sec:fits3}

In Sections \ref{sec:fits1} and \ref{sec:fits2}, we separately studied ${\bar B}\to D P$ and ${\bar B}\to {\bar D} P$ decays under hypotheses (a), the exact SU(3)$_F$ limit, and (b), including up to 30\% SU(3)$_F$ breaking. In both instances, we found that SU(3)$_F$ breaking was necessary, though overfitting was necessary due to the lack of data. In each set of decays, the amplitudes were proportional to a single weak phase, so that direct CP asymmetries were absent. While in some cases indirect CP asymmetries could have been included, they generally did not add any information (see Section \ref{Sec:observables}). We therefore focused entirely on branching ratios when analyzing ${\bar B}\to D P$ and ${\bar B}\to {\bar D} P$ decays separately.

In this section, we perform combined fits to the ${\bar B}\to D P$ and ${\bar B}\to {\bar D} P$ data using SU(3)$_F$ symmetry and its breaking. The SU(3)$_F$ formalism remains the same when the two sets of decays are combined: in the limit of exact SU(3)$_F$ symmetry, the amplitudes are described by seven diagrams, three for ${\bar B}\to D P$ decays and four for ${\bar B}\to{\bar D}P$ decays. Once SU(3)$_F$-breaking contributions are included, there are 18 relevant effective diagrams, eight from ${\bar B}\to D P$ and ten from ${\bar B}\to{\bar D}P$. The 15 branching ratios -- ten ${\bar B}\to D P$ and five ${\bar B}\to{\bar D}P$ -- are described in terms of these diagrams.

If this were the entire narrative, the combined fit would be uninteresting -- it could be broken into the two decoupled fits we had already performed. However, when the two sets of decays are combined, there are ten additional observables that are described by these same parameters. They are the six direct CP asymmetries listed in Table \ref{tab:observables_directCPV}, and the four indirect CP asymmetries listed in Table \ref{tab:observables_indirectCPV}. The addition of ten new observables makes these combined fits far more interesting. 

\begin{table}[!htbp]
\begin{center}
    \begin{tabular}{c|c} \hline
    SM weak phase  & World-average value \\ \hline
    $\beta$    &   $(22.2 \pm 0.7)^\circ$ \\
    $-2\,\beta_s$  &    $(-0.0368^{+0.0006}_{-0.0009})^{\circ}$ \\
    $\gamma$   &   $(66.2^{+3.4}_{-3.6})^{\circ}$ \\ 
    \hline
    \end{tabular}
\end{center}
\caption{World-average values of the SM weak phases used as external input in Fit 3 with hypotheses (a), (b), and (c). The value of $\beta$ is taken from Ref.~\cite{HFLAV:2019otj}, while those of $-2\,\beta_s$ and $\gamma$ are taken from Ref.~\cite{HFLAV:2022esi}.
\label{tab:ckmwa}}
\end{table}

Since CP asymmetries are included in these fits, the SM weak phases appear in the $\chi^2$ function. The CKM unitarity triangle angle $\gamma$ appears in the decay amplitudes, while $\beta_{(s)}$ appears in $B^0_{(s)}$-${\bar B}^0_{(s)}$ mixing. We treat these angles as external inputs using their world-average values given in Table \ref{tab:ckmwa}. In what follows, we once again test each hypothesis against this combined dataset of 25 observables.

\subsubsection{Hypothesis (a): exact SU(3)$_F$}
\label{sec:fits3a}

We perform a combined fit to all 25 observables in ${\bar B}\to DP$ and ${\bar B}\to{\bar D}P$ decays under hypothesis (a). The amplitudes are described by seven diagrams, which correspond to 13 real parameters to be fitted: seven magnitudes and six relative strong phases. This means that the number of d.o.f.~is 12. 

For this fit, we find $\chi^2_{\rm min}/{\rm d.o.f.} = 153.4/12$ with a corresponding $p$-value $\simeq 0$. Thus, hypothesis (a) is clearly rejected by the combined data. We present the best-fit parameter values in Table \ref{tab:DC12_SU(3)_fit}.

\begin{table}[h]
\begin{center}
    \begin{tabular}{c|c|c|c|c|c|c|c}
     \hline
        \multicolumn{4}{c|}{${\bar B}\to DP$ parameter} & \multicolumn{4}{c}{${\bar B}\to{\bar D}P$ parameter} \\ \hline
        Magnitude  & Best-fit value & Phase & Best-fit value & Magnitude  & Best-fit value & Phase & Best-fit value \\ \hline
        $|T|$ & $1.91 \pm 0.02 $ & -  &  - & $|T'|$ & $1.41\pm0.06$ & $\delta_{T'}$ & $(140\pm70)^\circ$ \\        
         $|C|$ & $0.77 \pm 0.04 $ & $\delta_C$  &  $(-75\pm 3)^\circ$ &  $|C'|$ & $0.7\pm0.2$ & $\delta_{C'}$ & $(130\pm30)^\circ$ \\
        $|E|$ & $0.19 \pm 0.01 $ & $\delta_E$  &  $(-160\pm10)^\circ$ &     $|E'|$ & $0.6\pm0.6$ & $\delta_{E'}$ & $(0\pm100)^\circ$ \\ 
        &&&& $|A'|$ & $0.1\pm0.1$ & $\delta_{A'}$ & $(-60\pm20)^\circ$ \\ \hline
\end{tabular}
\end{center}
\caption{Best-fit values of the fit parameters for the combined fit to ${\bar B}\to DP$ and ${\bar B}\to{\bar D}P$ decays, assuming hypothesis (a). The magnitudes of the diagrams are given in units of $10^{-5}$ GeV. The strong phase $\delta_T$ is taken to be zero.  
\label{tab:DC12_SU(3)_fit}}
\end{table}

\begin{table}[!htbp]
\begin{center}
    \begin{tabular}{c|c} \hline    
    Observable & $\chi^2_{\rm Obs}$  \\ \hline   
    $S(\bar{B^0_s}\to D_s^+K^-)$ & 5.1 \\
    $S(\bar{B^0}\to D^+\pi^-)$   & 8.7 \\
    $S(\bar{B^0}\to D^-\pi^+)$   & 6.8  \\
    $\mathcal{A}_{CP}(B^-\to D_{CP}\pi^-)$  & 23.8 \\ \hline
    \end{tabular}
\end{center}
\caption{Sizeable contributions to $\chi^2_{\rm min}$ from CP-violating observables in the combined fit to ${\bar B}\to DP$ and ${\bar B}\to{\bar D}P$ decays under hypothesis (a). $\chi^2_{\rm Obs}$ represents the contribution of each observable to the total $\chi^2_{\rm min}$. Only CP-violating observables with $\chi^2_{\rm Obs} > 1$ are included. 
\label{tab:chi23a}}
\end{table}

The rejection of hypothesis (a) by the combined data is not surprising. This is, to a great extent, driven by the branching ratios. Looking back at Secs.~\ref{sec:fits1a} and \ref{sec:fits2a}, we see that, in the SU(3)$_F$ limit, the total $\chi^2_{\rm min}$ contribution from separate fits to the branching ratios alone added up to $\sim 107$. This is still the case. Now that additional observables contribute to the fit, the $\chi^2_{\rm min}$ is even higher. In addition to the branching ratios, four observables give sizeable contributions to $\chi^2_{\rm min}$. In Table \ref{tab:chi23a}, we list these observables, along with their individual $\chi^2$ contributions. 

\subsubsection{Hypothesis (b): up to 30\% SU(3)$_F$ breaking}
\label{sec:fits3b}

The individual ${\bar B}\to DP$ and ${\bar B}\to{\bar D}P$ fits, which included only branching ratios, have found that, while hypothesis (a) is clearly rejected by the data, a reasonable agreement can be found with the addition of up to $\sim 30\%$ SU(3)$_F$ breaking, or hypothesis (b). Here we test the departure from SU(3)$_F$ symmetry by fitting to all 25 ${\bar B}\to DP$ and ${\bar B}\to{\bar D}P$ observables using an expanded set of 18 effective diagrams, or 35 real parameters, 18 magnitudes and 17 relative strong phases. As was shown in Secs.~\ref{sec:fits1b} and \ref{sec:fits2b}, these 18 diagrams consist of (i) seven diagrams that are the unbroken-SU(3)$_F$ diagrams with SU(3)$_F$-breaking corrections, and (ii) eleven combinations of SU(3)$_F$-breaking diagrams. The 18 diagram magnitudes are written as seven magnitudes of diagrams of type (i), and eleven ratios of the form (ii)/(i). We impose a constraint on all ratio parameters: $r_{X^{(\prime)}_i} \leq 30\%$ to test hypothesis (b).

Even though the number of fit parameters (35) far exceeds the number of observables (25), the result is a poor fit with $\chi^2_{\rm min} = 16.99$. The best-fit parameter values are given in Table \ref{tab:fit3b}. Although this fit is poor, the significant reduction in $\chi^2_{\rm min}$ implies that hypothesis (b) is far more preferred over hypothesis (a). Note that the best-fit values of all ratio parameters, $r_{X^{(\prime)}_i}$, saturate the imposed upper bound of 30\% in this fit. Together with the large $\chi^2_{\rm min}$, this demonstrates that the combined dataset prefers greater than 30\% SU(3)$_F$ breaking.

\begin{table}[!htbp]
\begin{center}
    \begin{tabular}{c|c|c|c|c|c|c|c}
     \hline
        \multicolumn{4}{c|}{${\bar B}\to DP$ parameter} & \multicolumn{4}{c}{${\bar B}\to{\bar D}P$ parameter} \\ \hline
        Magnitude  & Best-fit value & Phase & Best-fit value & Magnitude  & Best-fit value & Phase & Best-fit value \\ \hline
        $|\tilde{T}|$ & $1.83 \pm 0.04 $ & -  &  - & $|\tilde{T}'|$ & $1.9\pm0.2$ & $\delta_{\tilde{T}'}$ & $(-10\pm20)^\circ$ \\
        $|\tilde{C}|$ & $0.80 \pm 0.05 $ & $\delta_{\tilde{C}}$  &  $(75\pm 4)^\circ$ & $|\tilde{C}'|$ & $0.6\pm0.1$ & $\delta_{\tilde{C}'}$ & $(-170\pm10)^\circ$\\
        $|\tilde{E}|$ & $0.18 \pm 0.02 $ & $\delta_{\tilde{E}}$  &  $(150\pm 20)^\circ$ & $|\tilde{E}'|$ & $1.0\pm0.6$ & $\delta_{\tilde{E}'}$ & $(-160\pm30)^\circ$ \\
        &&&& $|\tilde{A}'|$ & $0.10\pm0.08$ & $\delta_{\tilde{A}'}$ & $(10\pm70)^\circ$ \\
        $r_{T_1}$ & $0.30 \pm 0.02 $ & $\delta_{T_1}$  &  $(71\pm 7)^\circ$ & $r_{T'_1}$ & $0.30\pm0.07$ & $\delta_{T'_1}$ & $(-150\pm50)^\circ$ \\
        $r_{T_2}$ & $0.3 \pm 0.3 $ & $\delta_{T_2}$  &  $(98\pm 5)^\circ$ & $r_{T'_2}$ & $0.3\pm0.2$ & $\delta_{T'_2}$ & $(-140\pm50)^\circ$ \\
        $r_{C_1}$ & $0.30 \pm 0.06 $ & $\delta_{C_1}$  &  $(25 \pm 20)^\circ$ & $r_{C'_1}$ & $0.30\pm0.03$ & $\delta_{C'_1}$ & $(260\pm20)^\circ$ \\
        $r_{C_2}$ & $0.3 \pm 0.3 $ & $\delta_{C_2}$  &  $(-300\pm 300)^\circ$ & &&& \\
        $r_{E_1}$ & $0.3 \pm 0.3 $ & $\delta_{E_1}$  &  $(100\pm 400)^\circ$ & $r_{E'_1}$ & $0.3\pm0.2$ & $\delta_{E'_1}$ & $(10\pm70)^\circ$ \\
        &&&& $r_{E'_2}$ & $0.3\pm0.2$ & $\delta_{E'_2}$ & $(10\pm70)^\circ$ \\
        &&&& $r_{A'_1}$ & $0.3\pm0.2$ & $\delta_{A'_1}$ & $(-190\pm80)^\circ$ \\ \hline
    \end{tabular}
\end{center}
\caption{Best-fit values of the 35 fit parameters for the combined fit to ${\bar B}\to DP$ and ${\bar B}\to{\bar D}P$ decays, assuming hypothesis (b). The magnitudes of the diagrams are given in units of $10^{-5}$ GeV. The strong phase $\delta_{\tilde{T}}$ is taken to be zero.
\label{tab:fit3b}}
\end{table}

It is worthwhile comparing these fit results with those of the individual fits assuming hypothesis (b) discussed in Secs.~\ref{sec:fits1b} and \ref{sec:fits2b}. In these earlier fits, we found that at most 30\% SU(3)$_F$ breaking was needed to explain the branching ratio data in ${\bar B}\to DP$ and ${\bar B}\to{\bar D}P$. Here we find that, once CP-violating observables are included in a combined fit, 30\% SU(3)$_F$ breaking is no longer sufficient. Apart from a deviation from isospin-symmetry expectations in two ${\bar B}\to{\bar D}P$ branching ratios (see the discussion in Sec.~\ref{sec:fits2b}), the dominant contributions to $\chi^2_{\rm min}$ come from two CP-violating observables. These observables are listed in Table \ref{tab:chi23b} along with their individual contributions to $\chi^2_{\rm min}$. 

\begin{table}[!htbp]
\begin{center}
    \begin{tabular}{c|c} \hline    
    Observable & $\chi^2_{\rm Obs}$  \\ \hline
    $S(\bar{B^0}\to D^-\pi^+)$   & 2.8  \\
    $\mathcal{A}_{CP}(B^-\to D_{CP}\pi^-)$  & 10.3 \\ \hline
    \end{tabular}
\end{center}
\caption{Sizeable contributions to $\chi^2_{\rm min}$ from CP-violating observables in the combined fit to ${\bar B}\to DP$ and ${\bar B}\to{\bar D}P$ decays under hypothesis (b). \label{tab:chi23b}}
\end{table}

\subsubsection{Hypothesis (c): up to 60\% SU(3)$_F$ breaking}
\label{sec:fits3c}

In Sec.~\ref{sec:fits3b}, we found that the results of the fit to the combined dataset for ${\bar B}\to DP$ and ${\bar B}\to{\bar D}P$ decays, which includes branching ratios and CP asymmetries, indicate that larger than 30\% SU(3)$_F$ breaking is necessary to explain the data. We now perform a $\chi^2$ minimization fit allowing up to 60\% SU(3)$_F$ breaking, i.e., we have the same unknown parameters as in Sec.~\ref{sec:fits3b}, but impose a larger upper bound: $r_{X^{(\prime)}_i} \leq R = 60\%$.

Allowing for 60\% SU(3)$_F$ breaking, we fit the 25 observables that depend on 35 parameters. We find, $\chi^2_{\rm min} = 2.8$. The best-fit values of the 35 parameters are given in Table \ref{tab:fit3c}. There are two main contributions to this $\chi^2_{\rm min}$. One of these contributions was discussed previously in Sec.~\ref{sec:fits2b}: there is a pair of isospin-symmetry-related branching ratios whose measured values deviate from the isospin relation by $\sim 1\sigma$, adding $\sim 1.0$ to the minimum $\chi^2$. 

\begin{table}[!htbp]
\begin{center}
    \begin{tabular}{c|c|c|c|c|c|c|c}
     \hline
        \multicolumn{4}{c|}{${\bar B}\to DP$ parameter} & \multicolumn{4}{c}{${\bar B}\to{\bar D}P$ parameter} \\ \hline
        Magnitude  & Best-fit value & Phase & Best-fit value & Magnitude  & Best-fit value & Phase & Best-fit value \\ \hline
        $|\tilde{T}|$ & $1.79 \pm 0.03 $ & -  &  - & $|\tilde{T}'|$ & $2.0\pm0.2$ & $\delta_{\tilde{T}'}$ & $(20\pm10)^\circ$ \\
        $|\tilde{C}|$ & $0.84 \pm 0.03 $ & $\delta_{\tilde{C}}$  &  $(-74\pm 3)^\circ$ & $|\tilde{C}'|$ & $1.1\pm0.1$ & $\delta_{\tilde{C}'}$ & $(-44\pm4)^\circ$ \\
        $|\tilde{E}|$ & $0.18 \pm 0.02 $ & $\delta_{\tilde{E}}$  &  $(-140\pm10)^\circ$ & $|\tilde{E}'|$ & $1.0\pm0.8$ & $\delta_{\tilde{E}'}$ & $(-140\pm20)^\circ$ \\
        &&&& $|\tilde{A}'|$ & $0.10\pm0.08$ & $\delta_{\tilde{A}'}$ & $(140\pm40)^\circ$ \\
        $r_{T_1}$ & $0.6 \pm 0.5 $ & $\delta_{T_1}$  &  $(-90\pm 5)^\circ$ & $r_{T'_1}$ & $0.3\pm0.1$ & $\delta_{T'_1}$ & $(-200\pm300)^\circ$ \\
        $r_{T_2}$ & $0.1 \pm 0.4 $ & $\delta_{T_2}$  &  $(80\pm 40)^\circ$ & $r_{T'_2}$ & $0.6\pm0.3$ & $\delta_{T'_2}$ & $(100\pm90)^\circ$ \\
        $r_{C_1}$ & $0.5 \pm 0.1 $ & $\delta_{C_1}$  &  $(-76\pm10)^\circ$ & $r_{C'_1}$ & $0.6\pm0.5$ & $\delta_{C'_1}$ & $(146\pm6)^\circ$ \\
        $r_{C_2}$ & $0.2 \pm 0.4 $ & $\delta_{C_2}$  &  $(200\pm200)^\circ$ & &&& \\
        $r_{E_1}$ & $0.6 \pm 0.3 $ & $\delta_{E_1}$  &  $(100\pm300)^\circ$ & $r_{E'_1}$ & $0.6\pm0.5$ & $\delta_{E'_1}$ & $(-100\pm300)^\circ$ \\
        &&&& $r_{E'_2}$ & $0.6\pm0.4$ & $\delta_{E'_2}$ & $(-100\pm200)^\circ$ \\
        &&&& $r_{A'_1}$ & $0.6\pm0.4$ & $\delta_{A'_1}$ & $(240\pm40)^\circ$ \\ \hline
    \end{tabular}
\end{center}
\caption{Best-fit values of the 35 fit parameters for the combined fit to ${\bar B}\to DP$ and ${\bar B}\to{\bar D}P$ decays, assuming hypothesis (c). The magnitudes of the diagrams are given in units of $10^{-5}$ GeV. The strong phase $\delta_{\tilde{T}}$ is taken to be zero. \label{tab:fit3c}}
\end{table}

We also find an additional contribution to $\chi^2_{\rm min}$ of $\sim 1.4$ that cannot be removed. This can only arise from a model-independent relationship between the indirect CP asymmetries and branching ratios in the decays ${\bar B}^0\to D^+\pi^-$ and ${\bar B}^0\to D^-\pi^+$. One possibility is the following. The indirect CP asymmetries in these two decays can be expressed as
\beq
  S({\bar B}^0\to D^\pm\pi^\mp) ~=~ -\,\frac{2\,\kappa}{1 + \kappa^2}~ \sin(2\beta+\gamma\mp\Delta) ~, 
\label{eq:spm}
\eeq
where $\kappa \equiv \sqrt{{\cal B}(\bar{B^0}\to D^+\pi^-)/{\cal B}(\bar{B^0}\to D^-\pi^+)}$ and $\Delta$ is the strong-phase difference between the decay amplitudes $A({\bar B}^0\to D^+\pi^-)$ and $A({\bar B}^0\to D^-\pi^+)$. Using ${\cal B}^{+-(-+)}$ and $S^{+-(-+)}$ to denote the branching ratio and indirect CP asymmetry for ${\bar B}^0\to D^+\pi^-({\bar B}^0\to D^-\pi^+)$, the above equation can be rewritten as follows:
\beq
\(\frac{S^{+-} + S^{-+}}{\sin(2\beta + \gamma)}\)^2 + 
\(\frac{S^{+-} - S^{-+}}{\cos(2\beta + \gamma)}\)^2 - \frac{16\,{\cal B}^{+-}{\cal B}^{-+}}{({\cal B}^{+-} + {\cal B}^{-+})^2} ~=~ 0\,.\label{eq:bdpirel}
\eeq
Using the measured values of ${\cal B}^{+-}$, ${\cal B}^{-+}$, $S^{+-}$, and $S^{-+}$, we find that the left-hand side of the above relationship yields $0.009\pm0.013$. This is consistent with zero, so this is not the source of the additional contribution of $\sim 1.4$ to $\chi^2_{\rm min}$.

In fact, this comes from a different relationship between the indirect CP asymmetries and branching ratios in the decays ${\bar B}^0\to D^+\pi^-$ and ${\bar B}^0\to D^-\pi^+$. In Eq.~(\ref{eq:spm}), we have $\kappa = 59 \pm 5$. The central value of the ratio $2\,\kappa/(1 + \kappa^2) = 0.0341\pm0.0029$ places an upper bound on the magnitudes of both indirect CP asymmetries. However, the measured central values of both asymmetries violate this bound. Therefore, it is not possible to extract $\Delta$ from Eq.~(\ref{eq:spm}). On the other hand, the asymmetries can be combined with $\kappa$ eliminated as follows:
\beq
\frac{S({\bar B}^0\to D^+\pi^-)}{S({\bar B}^0\to D^-\pi^+)} ~=~ \frac{\sin(2\beta+\gamma-\Delta)}{\sin(2\beta+\gamma+\Delta)} ~=~ 0.7\pm0.4 ~. 
\label{eq:sc}
\eeq
Taking $0 \leq \Delta \leq 180^\circ$ to avoid discrete ambiguities, we find that $\Delta = (155 \pm 28)^\circ$ satisfies Eq.~(\ref{eq:sc}). Using this value of $\Delta$ in the right-hand side of Eq.~(\ref{eq:spm}), we obtain 
\beq
S({\bar B}^0\to D^+\pi^-)_{\rm th} ~=~ 0.024 \pm 0.013\,,\qquad S({\bar B}^0\to D^-\pi^+)_{\rm th} ~=~ 0.0340 \pm 0.0032 ~,
\eeq
where the subscript`${\rm th}$' has been added to distinguish these ``theoretical'' indirect CP asymmetries from their measured values. Comparing these values with their respective measurements in Tables \ref{tab:Br_data_DC1} and \ref{tab:DC2_Br}, a rough estimate of their contribution to $\chi^2_{\rm min}$ that cannot be removed in a fit turns out to be $\chi^2 \sim 1.4$.

We now vary $R$ from 0 to 60\% and plot $\chi^2_{\rm min}$ as a function of $R$ in Fig.~\ref{fig:chi2Rf3}. As before, $R = 0$ represents exact SU(3)$_F$ symmetry: we find $\chi^2_{\rm min} = 153.4$. For $R \geq 60\%$, i.e.~when greater than 60\% SU(3)$_F$ breaking is allowed, we find $\chi^2_{\rm min} = 2.8$. Thus, while 30\% SU(3)$_F$ breaking is not sufficient to explain the combined data in ${\bar B}\to DP$ and ${\bar B}\to{\bar D}P$, we can conclude that 60\% SU(3)$_F$ breaking {\it is} sufficient.

\begin{figure}[!htbp]
\centering
\includegraphics[width=0.75\textwidth]{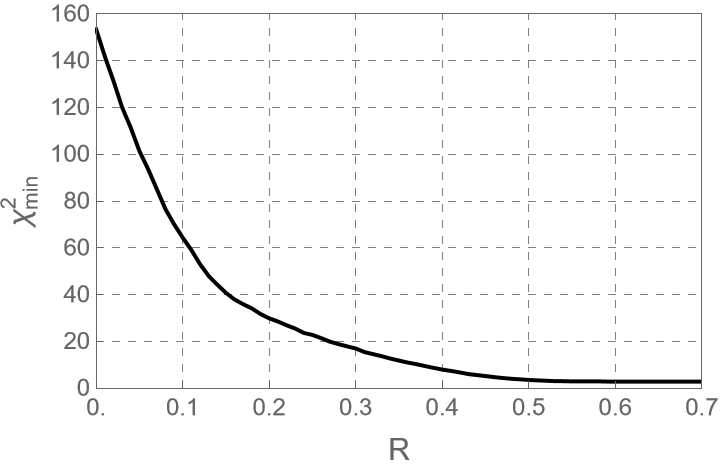}
\caption{$\chi^2_{\rm min}$ as a function of $R$ for fits to ${\bar B}\to DP$ and ${\bar B}\to{\bar D}P$ data. Here $R$ is the upper limit set on all ratios $r_{ X^{(\prime)}_i}$ in a given fit. There is a residual $\chi^2_{\rm min} \sim 2.8$ for $R \gtrsim 60\%$ that cannot be removed by adding further SU(3)$_F$ breaking.\label{fig:chi2Rf3}}
\end{figure}

\section{Conclusions}

Recently, in Refs.~\cite{Berthiaume:2023kmp, Bhattacharya:2025wcq}, $B\to PP$ decays ($B = \{B^+, B^0, B_s^0\}$, $P \in \{ \pi, K, \eta, \eta' \}$) were examined under the assumption of SU(3)$_F$, flavor SU(3) symmetry. Fits to the latest data revealed a $\sim 4\sigma$ disagreement with the SU(3)$_F$ limit of the SM, SM$_{\rm{SU(3)}_F}$. The addition of 1000\% SU(3)$_F$-breaking effects is required to remove this discrepancy, which is considerably larger than the $\sim 30\%$ breaking expected in the SM. Following this result, ${\bar B} \to D P$ decays ($D = \{D^0, D^+, D_s^+\}$, $P \in \{ \pi, K\}$) were analyzed in Ref.~\cite{Davies:2024vmv}, again assuming SU(3)$_F$. An even larger disagreement with SM$_{\rm{SU(3)}_F}$ was found, $> 5\sigma$. However, here it was shown that this discrepancy can be removed by adding $\sim 20\%$ SU(3)$_F$-breaking effects.

The only observables used in this analysis were the branching ratios of ${\bar B} \to D P$ decays. However, the ${\bar B} \to D^0 P$ amplitudes can contribute to other observables: they can interfere with the amplitudes for ${\bar B} \to {\bar D}^0 P$ decays if the $D^0$ and ${\bar D}^0$ decay to the same final state. There are several final states for which this can occur, and this interference can lead to direct and indirect CP asymmetries.

In this paper, we have extended the analysis of Ref.~\cite{Davies:2024vmv} to include both ${\bar B} \to D P$ and ${\bar B} \to  \bar D P$ decays. We find the following results:
\begin{enumerate}[label=(\roman*)]

\item We have updated the analysis of Ref.~\cite{Davies:2024vmv}. We analyze ${\bar B} \to D P$ decays, including only the branching ratios as observables. In the limit of exact SU(3)$_F$, we again find a disagreement of $> 5\sigma$ with SM$_{\rm{SU(3)}_F}$. 

In light of this, we separately analyze $\Delta S=0$ and $\Delta S=1$ ${\bar B} \to D P$ decays. In both cases, the data can be explained by a single set of parameters. A comparison of the best-fit values of these parameters finds that they differ by 20-30\%.

We then explicitly add SU(3)$_F$-breaking effects, and find that the $> 5\sigma$ discrepancy can be removed by adding $\sim 20\%$ breaking.

\item We analyze ${\bar B} \to \bar D P$ decays, using only branching ratios. Here things are more complicated: there are more theoretical parameters than observables, so overfitting is necessary. But we do not find $\chi^2_{\rm min} = 0$. This is due to two discrepancies in the data. There are two amplitudes related by isospin and two others related by U spin, for which the data do not respect these relations.

Of these two discrepancies, only the second can be improved by adding SU(3)$_F$-breaking effects. We find that, when 10\% breaking is added, the U-spin discrepancy is removed, leaving a value of $\chi^2_{\rm min}$ consistent with the isospin discrepancy.

\item We perform a combined analysis of ${\bar B} \to D P$ and ${\bar B} \to \bar D P$ decays, using all the data, 15 branching ratios and ten direct and indirect CP asymmetries. As was the case in the ${\bar B} \to D P$ analysis, we find a disagreement of $> 5\sigma$ with SM$_{\rm{SU(3)}_F}$ in the limit of exact SU(3)$_F$ symmetry. 

Adding SU(3)$_F$-breaking effects, we find that the $> 5\sigma$ discrepancy can be removed (except for the isospin discrepancy) with $\sim 60\%$ breaking.

\end{enumerate}

\acknowledgments{We thank Keri Vos for pointing out typographical errors in the original manuscript. This work was financially supported by the National Science Foundation under Grant No.\ PHY-2609773 (B.B.) and by NSERC of Canada (R.B., D.L.).}

\appendix

\section{Equivalence of RMEs and Diagrams}
\label{RME_Diag_Equiv}

\subsection{The SU(3)$_F$-invariant Case}

In Section \ref{sec:formalism}, we expressed the amplitudes for ${\bar B}\to DP$ and ${\bar B}\to{\bar D}P$ decays in terms of both SU(3)$_F$ reduced matrix elements (RMEs) and topological diagrams. For each set of decays, the RMEs can be expressed as linear combinations of the diagrams. For the ${\bar B}\to DP$ decays, three RMEs are related to three diagrams as follows. 
\beq\label{eq:diag_RMEs}
O_{\bar{15}} = T + C ~,~~
O_{6} = C - T ~,~~ 
O_{\bar{3}} = -\frac{3\,T - C + 8\,E}{3}~.~~
\eeq
For ${\bar B}\to {\bar D}P$, four RMEs can be expressed in terms of four diagrams as follows.
\beq
S_{15} = \frac{T + C}{\s}~,~~
T_{\bar{6}} = \frac{T - C}{\s}~,~~
S_3 = \frac{T + C + 4(E + A)}{\sqrt{6}}~,~~
T_3 = \frac{C - T + 2(A - E)}{\sqrt{3}}~.~~
\eeq
The above relationships demonstrate a one-to-one correspondence between SU(3)$_F$ RMEs and topological diagrams in the SU(3)$_F$-symmetry limit. Therefore, in this limit, topological diagrams are equivalent to SU(3)$_F$ RMEs. 

\subsection{The SU(3)$_F$-breaking Case}

In ${\bar B}\to DP$ decays, once first-order SU(3)-breaking corrections are included, we find eight RMEs and eight diagrams. These RMEs and diagrams are related as follows.
\begin{eqnarray}\label{eq:Diag_RME_relations_DC1}
    \tilde{O}_{\bar{15}} &=& (\tilde{T} + \tilde{C}) + \frac{2\,(\tilde{T}_1 + \tilde{C}_1)}{3} + \frac{\tilde{T}_2 + \tilde{C}_2}{9}\,, \nl
    \tilde{O}_{6} &=& (\tilde{C}-\tilde{T}) + \frac{2\,(\tilde{C}_1 - \tilde{T}_1)}{3} + \frac{3\,\tilde{C}_2 - 5\,\tilde{T}_2}{9} \,, \nl
    \tilde{O}_{\bar{3}} &=& -\frac{3\,\tilde{T} - \tilde{C} + 8\,\tilde{E}}{3} - \frac{6\,\tilde{T}_1 + 9\,\tilde{T}_2 - 2\,\tilde{C}_1 + \tilde{C}_2 + 16\,\tilde{E}_1}{9}\,, \\
    \tilde{O}^1_{\bar{15}} &=& \frac{2\,(\tilde{T}_1 + \tilde{C}_1)}{3} - \frac{2\,(\tilde{T}_2 + \tilde{C}_2)}{9}\,,~~~\tilde{O}^1_{\bar{3}} ~=~ \frac{6\,\tilde{T}_1 + 6\,\tilde{T}_2 - 2\,\tilde{C}_1 - 2\,\tilde{C}_2 + 16\,\tilde{E}_1}{9}\,, \nl
    \tilde{O}^1_6 &=& \frac{6\,\tilde{T}_1 + 2\,\tilde{T}_2 - 6\,\tilde{C}_1 + 6\,\tilde{C}_2}{9}\,,~~~\tilde{W}_{\bar{15}} ~=~ -\frac{2\,(\tilde{T}_2 + \tilde{C}_2)}{3}\,,~~~ \tilde{V}_6 ~=~ \frac{\sqrt{10}\,\tilde{T}_2}{3}\,. \nn
\end{eqnarray}
The above relationships demonstrate the equivalence between diagrams and RMEs for ${\bar B}\to DP$ decays including first-order SU(3)$_F$ breaking.

In ${\bar B}\to {\bar D}P$ decays, once SU(3)$_F$ breaking is included, we find 11 RMEs and 11 diagrams. The following relationships establish the equivalence between these RMEs and diagrams. 
\begin{eqnarray}\label{eq:Diag_RME_relations_DC2}
    \tilde{S}_{15} &=& \frac{1}{153\sqrt{2}} \left[ 139\,\tilde{T} - 37\,\tilde{C} - 48\left(\tilde{E}-\tilde{A}\right) + 146\,\tilde{T}_1 + 23\,\tilde{T}_2\right. \nl
    &&\hspace{4truecm} \left.+~58\,\tilde{C}_1 + 63\,\tilde{C}_2 + 24\left( \frac{7}{3}\,\tilde{E}_2 - \tilde{E}_1 + \tilde{A}_1 \right)\right], \nl
    \tilde{T}_{\bar{6}} &=& \frac{1}{102\sqrt{2}} \left[ -14 \left( \tilde{T} - \tilde{C} \right) - 48\left( \tilde{E}- \tilde{A} \right) + 44\left( \tilde{T}_1 - \tilde{C}_1 \right) \right. \nl
    &&\hspace{45truemm} \left. +~ 23\,\tilde{T}_2 - 39\,\tilde{C}_2 - 24\left( \tilde{E}_1 - \tilde{A}_1 \right) + 5\,\tilde{E}_2  \right], \nl
    \tilde{S}_3 &=& \frac{1}{51\sqrt{6}} \left[ 9\,\tilde{T} + 25\,\tilde{C} + 196\,\tilde{E} - 60\,\tilde{A} + 30\,\tilde{T}_1 + 69\,\tilde{T}_2 \right. \nl
    &&\hspace{4truecm} \left. +~ 38\,\tilde{C}_1 - 49\,\tilde{C}_2 + 200\,\tilde{E}_1 + 66\,\tilde{E}_2 + 72\,\tilde{A}_1 \right], \nl
    \tilde{T}_3 &=& \frac{1}{51\sqrt{3}} \left[ 95\left(\,\tilde{T} -\,\tilde{C} \right) + 78\left( \tilde{E} - \,\tilde{A} \right) + 22 \left( \,\tilde{T}_1 - \tilde{C}_1 \right)\right. \nl
    &&\hspace{45truemm} \left. -~ 14\,\tilde{T}_2 + 6\,\tilde{C}_2 + 12\left( \tilde{A}_1 - \tilde{E}_1 \right) - 23\,\tilde{E}_2  \right], \nl
    \tilde{M}^1_{15} &=& -\frac{2}{17\sqrt{3}} \left[ 22\left( \tilde{T}-\tilde{C} \right) - 12\left( \tilde{E} - \tilde{A} \right) + 11\left( \tilde{T}_1 - \tilde{C}_1 \right)\right. \nl
    &&\hspace{5truecm} \left. -~ 7\,\tilde{T}_2 + 3\,\tilde{C}_2 - 6\left( \tilde{E}_1 - \tilde{A}_1 \right) + 14\,\tilde{E}_2 \right], \nl
    \tilde{M}^2_{15} &=& -\frac{\sqrt{6}}{17} \left[ 8\left( \tilde{T} - \tilde{C} + \tilde{E} - \tilde{A} \right) + 4\left( \tilde{T}_1 - \tilde{C}_1 + \tilde{E}_1 - \tilde{A}_1 \right) \right.   \nl
    &&\hspace{6truecm} \left. -~ \left(\tilde{T}_2 + 2\,\tilde{C}_2 - 2\,\tilde{E}_2 \right) \right], \\
    \tilde{M}^1_{\bar{6}} &=& -\frac{\sqrt{5}}{102\,\s}\left[ 88\left( \tilde{T} - \tilde{C} \right) - 48\left( \tilde{E} - \tilde{A} \right) + 44\left( \tilde{T}_1 - \tilde{C}_1 \right)  \right. \nl 
    &&\hspace{4truecm} \left. -~79\,\tilde{T}_2 + 63\,\tilde{C}_2 - 24\left( \tilde{E}_1 - \tilde{A}_1 \right) + 5\,\tilde{E}_2 \right], \nl
    \tilde{M}^2_{\bar{6}} &=& \frac{\sqrt{5}}{34} \left[ 48\left( \tilde{T} - \tilde{C} + \tilde{E} - \tilde{A} \right) + 24\left( \tilde{T}_1 - \tilde{C}_1 + \tilde{E}_1 - \tilde{A}_1 \right)\right. \nl
    &&\hspace{6truecm} \left. +~ 11\,\tilde{T}_2 + 5\left(\tilde{C}_2 - \tilde{E}_2 \right) \right], \nl
    \tilde{R}_{15} &=& \frac{4\sqrt{3}}{85} \left[ 32\left( \tilde{T} - \tilde{C} + \tilde{E} - \tilde{A} \right) + 16\left( \tilde{T}_1 - \tilde{C}_1 + \tilde{E}_1 - \tilde{A}_1 \right) \right.\nl 
    && \hspace{6truecm} \left.+~ 13\,\tilde{T}_2 + 9\,\tilde{C}_2  + 8\,\tilde{E}_2 \right], \nl
    \tilde{S}^1_{15} &=& \frac{\sqrt{2}}{765} \left[ 946\,\tilde{T} + 74\,\tilde{C} + 96\left( \tilde{E} - \tilde{A} \right) + 473\,\tilde{T}_1 - 46\,\tilde{T}_2 \right. \nl
    &&\hspace{4truecm} \left. +~ 37\,\tilde{C}_1 + 27\,\tilde{C}_2 + 48\left( \tilde{E}_1 - \tilde{A}_1 \right) + 194\,\tilde{E}_2 \right], \nl
    \tilde{S}^1_{3} &=& -\sqrt{\frac{2}{3}}\left[2\left(-\tilde{T} + \frac{\tilde{C}}{3} - \frac{8\,\tilde{E}}{3}\right) - \tilde{T}_1 - \tilde{T}_2 + \frac{\tilde{C}_1 + \tilde{C}_2}{3} - \frac{8\,\tilde{E}_1}{3}\right]\,. \nn
\end{eqnarray}

\bibliography{B2DP}
\bibliographystyle{apsrev4-2-titles}

\end{document}